\documentclass[aps,prb,twocolumn,superscriptaddress,floatfix,amsmath,onlytext,fullfamily,textlf,longbibliography]{revtex4-2} 

\usepackage[normalem]{ulem}
\usepackage{bm} 
\usepackage{amssymb}
\usepackage{amsfonts} 
\usepackage{amsmath} 

\usepackage{graphicx} 

\usepackage{xcolor} 
 
\usepackage[pdftex,colorlinks=true,allcolors=blue,breaklinks=true]{hyperref}  

\usepackage[utf8]{inputenc} 
\usepackage{textgreek}
 
\definecolor{g-blue}{rgb}{0.83,0.95,1}
\definecolor{g-yellow}{rgb}{1,1,0.7}
\definecolor{g-green}{rgb}{0.9,1,0.9}
\definecolor{green}{rgb}{0,0.6,0}
\definecolor{cyan}{rgb}{0,0.7,0.7}
\definecolor{black}{rgb}{0,0,0}
\definecolor{grey}{rgb}{0.4,0.4,0.4}
\definecolor{nature-blue}{rgb}{0.0,0.200,0.500}

\def \ed {\end{document}}
\def\Fbox#1{\vskip1ex\hbox to 8.5cm{\hfil\fboxsep0.3cm\fbox{%
		\parbox{8.0cm}{#1}}\hfil}\vskip1ex\noindent}  

\def\be{\begin{equation}}
\def\ee{\end{equation}}
\def\bea{\begin{eqnarray}}
\def\eea{\end{eqnarray}}
\def\bse{\begin{subequations}}
\def\ese{\end{subequations}}

\let \= \equiv  
\let\*\cdot 
\let\^\widehat 
\let\-\overline

\def\1{\bm1}

\def\<{\left\langle}    \def\>{\right\rangle}
\def\({\left(}          \def\){\right)}
\def\[ {\left[}         \def\]{\right]}

\newcommand{\B}[1]{{\bm{#1}}}

\renewcommand{\sb}[1]{_{\text {#1}}}  
\renewcommand{\sp}[1]{^{\text {#1}}}  
\def\Sb#1{_{\scriptscriptstyle\rm{#1}}}
 
\begin{document}

	\title{ Precritical  anomalous scaling and  magnetization temperature dependence \\  in cubic ferromagnetic crystals}
 \author{Igor Kolokolov}
 \email{kolokol@itp.ac.ru}
 \affiliation{L.D. Landau Institute for Theoretical Physics, Ak. Semenova 1-A, Chernogolovka 142432, Moscow region, Russia}
 \affiliation{ National Research University  Higher School of Economics, 
 101000, Myasnitskaya 20, Moscow, Russia}
	\author{Victor~S.~L'vov}
 	\email{victor.lvov@gmail.com}	
    \affiliation{Department of Physics of Complex Systems, Weizmann Institute of Science, Rehovot 76100, Israel}
	
    \author{Anna Pomyalov}
 	\email{anna.pomyalov@weizmann.ac.il}	
	\affiliation{Department of Chemical and Biological Physics, Weizmann Institute of Science, Rehovot 76100, Israel}

\begin{abstract} 
 Recent developments in spintronics have drawn renewed attention to the spin dynamics of cubic ferromagnetic crystals EuO and EuS.
These ferromagnets have the simplest possible magnetic structure, making them the most suitable systems for testing various theoretical models of magnetic materials. 
   A commonly used Weiss mean-field approximation  (MFA) 
   provides only a qualitative description of the magnetization temperature dependence $M(T)$. 
   We develop a consistent theory for $M(T)$, based on the perturbation diagrammatic technique for spin operators.
 Our theory is in excellent quantitative agreement with the experimental dependence of $M(T)$ for EuO and EuS throughout the entire temperature range from $T=0$ to Curie temperature $T\Sb{C}$. In particular, our theoretical dependence $M(T)$ demonstrates a scaling behavior $M(T)\propto (T\Sb {C}-T)^{\beta_*}$ with the scaling index $\beta_*\approx 1/3$ in a wide range of temperatures in agreement with the experimentally observed apparent scaling in EuO and EuS. 
The scaling behavior with $\beta_*\approx 1/3$ is manifested in the temperature range $T\lesssim  T\Sb{C}$ 
 where corrections to the magnetization  due to its fluctuations   $\delta M (T)\lesssim  M(T)$. To distinguish it from the narrow ``critical"  range $T\approx T\Sb{C}$ with $\delta M (T) > M(T)$, we term this $T$-range ``precritical". The precritical corrections $\delta M (T)$  are still large enough to affect the $M(T)$ behavior. 
The index $\beta_*$ fundamentally differs from the ``normal" scaling index $\beta\Sb{MFA}=1/2$ predicted by the MFA, which neglects the magnetization fluctuations.  We refer to the emerging in our theory apparent magnetization scaling with  $\beta_*\approx 1/3$  as ``precritical anomalous".
\end{abstract}

 \maketitle 
\numberwithin{equation}{section}


  \section{\label{s:Intro} Introduction}

\subsection{\label{ss:EuO} Ferromagnetics EuO and EuS}

The development of spintronics has generated significant interest in rare-earth oxide ferromagnetic semiconductors, such as europium chalcogenides EuO and EuS.
In particular,  EuO is especially promising for applications 
 \cite{Mairoser2010} as it has the third strongest saturation magnetization of all known ferromagnets \cite{Matthias1961}, one of the largest magneto-optic Kerr effects\,\cite{Ahn1967}, a pronounced
insulator-to-metal transition\,\cite{Shafer1972,Burg2012,Sinjukow2003}   as well as a colossal
magnetoresistance effect\,\cite{Shapira1973}.

\begin{figure} 
  \includegraphics[width=0.8\columnwidth]  {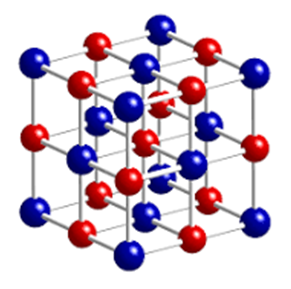}  
        \caption{\label{f:1} FCC crystallographic structure of EuO and EuS. Blue balls represent Eu atoms, and red balls represent oxygen (O)  or sulfur (S) atoms. The lattice constant is
$a=5.14\,$\AA \ for EuO and  $a=5.96\,$ \AA \ for EuS.   Curie temperatures are $T\Sb C=69.2\,$K for EuO and  $T\Sb C=16.6\,$K for EuS. In both structures, the coordination number of the Eu atoms is $Z=12$. For additional parameters, see Tab.\,\ref{t:1}.
           }
\end{figure}

Moreover, among various magnetically ordered materials, EuO and EuS are probably the most suitable systems for testing various theoretical models of magnetic material because they are well-studied experimentally and have a simple crystallographic structure. Unlike other well-known magnetics, such as Yttrium Iron Garnet (YIG), which has 80 atoms in the unit cell with 20 of them (Fe) possessing a magnetic moment\,\cite{Cherepanov1993}, EuO and EuS are the only known ferromagnets having two atoms in the unit cell, with only one of them  (Eu$^{+2}$, with  spin $S=7/2$)  having a magnetic moment. The crystallographic structure of EuO and EuS [face-centered cubic (FCC) lattice] is illustrated in Fig.~\ref{f:1}.   

\begin{figure*} 
\begin{tabular}{cc}
   \hskip -.5 cm   \includegraphics[width=1.15  \columnwidth]{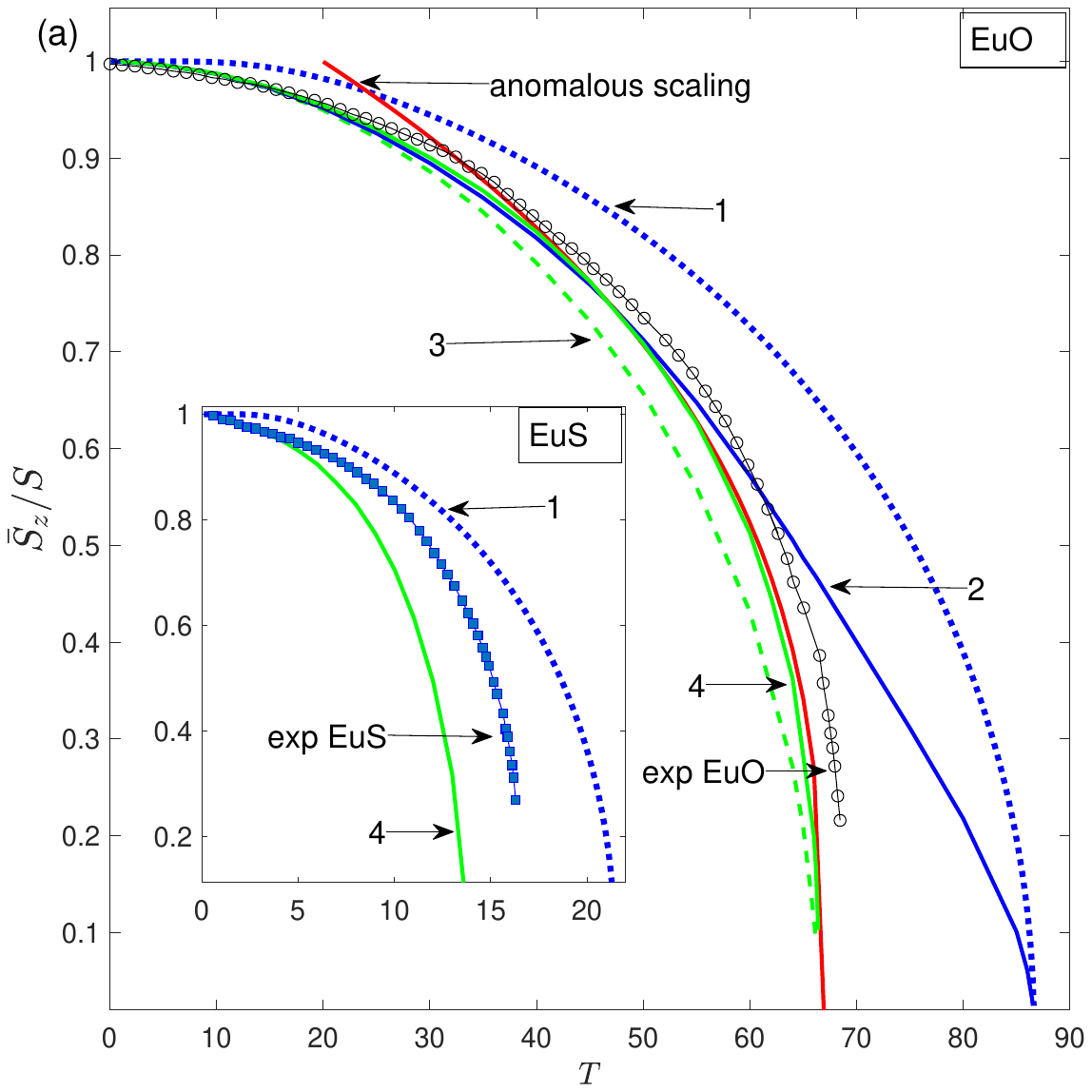} &
\hskip -.9 cm   \includegraphics[width=1.16  \columnwidth]{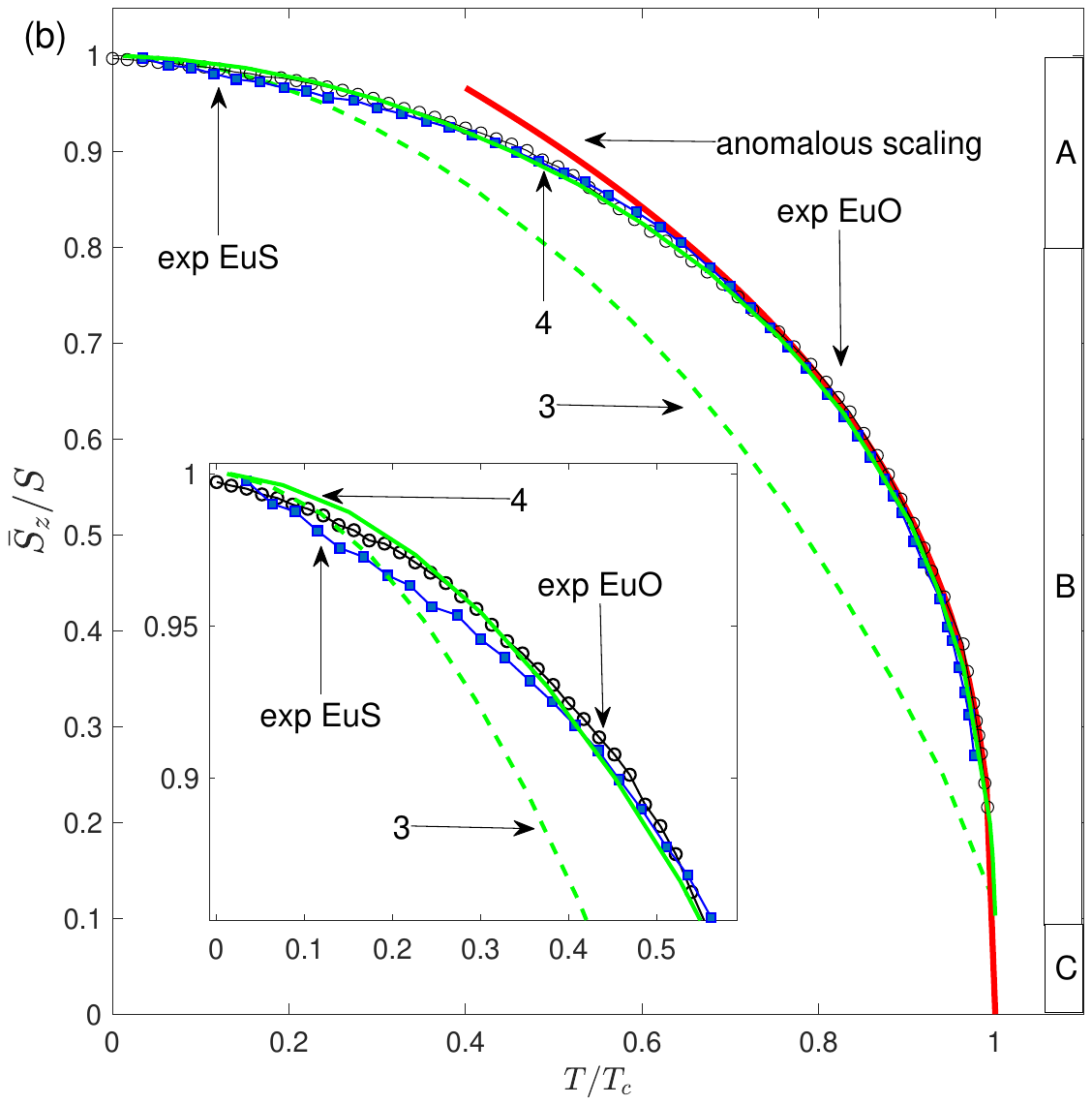} 
  \end{tabular}
 \caption{\label{f:2}  
  Experimental and numerical results  for $\overline{S}_z/S$ vs. $T$ [panel (a)] and vs. $T/T_{_{\rm C}}$ [panel (b)]. In panel (a), the results for EuO are summarized in the main panel, and those for EuS are summarized in the inset. The inset in (b) gives a close-up of the low-$T$ range for the same data as in the main panel of (b).  Experimental results\,\cite{Passell1976,Als-Nielsen1976,Mook1981} for EuO are plotted as solid black lines with circles, and those for  EuS are plotted by solid blue lines with squares.   The results of numerical solutions are denoted as follows:\\
(1)  Quantum version \eqref{QWEa} of the Weiss-Heisenberg  MFA for EuO in the main panel  and for EuS in the inset of (a)
  --dotted blue lines;\\
(2) $1/ Z$-corrected equation of the Weiss-Heisenberg MFA \,\eqref{I1.22A} for EuO ---solid  blue line.  Recall that  $Z=12$ is the coordination number of Eu atoms, see Fig.\,\ref{f:1}.\\ 
(3) Spin-wave-improved version of the MFA  \eqref{1.20}   for EuO-- dashed 
 green  line;\\
 (4) Spin-wave-improved version with the first-order $1/Z$-  corrections  \eqref{eq5}  for EuO in the main panel of (a) and (b),  and for EuS in the inset of (a) --- solid  green  line;\\   
 The  red  line represents the power-law fit to the solution of Eq.\,\eqref{eq5} in the form 
 $a_{\rm th} ( T_{_{\rm C}}- T)^{\beta_*}$, $\beta_*= 0.34 \pm 0.02$
 with   $T_{_{\rm C}}=66.5$.   The  predicted  coefficients are $a_{\rm th}=0.27$ in (a) and $a_{\rm th}=1.15$ in (b). \\
Three regimes of $\overline S_z(T)/S$ behavior dominated by various types of magnetization fluctuations are marked in (b). In the low-temperature ``spin-wave dominated" regime A, the $T$-dependence of $\overline S_z(T)$ is determined by the excitation level of spin waves (the transversal fluctuations of spins). In the precritical regime B, moderate transversal and longitudinal fluctuations are equally important, leading to deviation from MFA predictions. In the critical regime C, the corrections to the mean spin exceed its mean value, and the MFA, along with all our corrections, is no longer valid. Other approaches (e.g., the renormalization group approach \cite{Guillou1977}) are required in this regime.   }
\end{figure*}

\subsection{ Plan of the paper and main results} 
   This paper aims to describe and improve a theory of spontaneous magnetization $M(T)$   of ferromagnetics over an entire temperature range from $T=0$ to Curie temperature $T_{_{\rm C}}$, at which $M=0$. The theory is evaluated by comparing it with the existing experimental data \cite{Passell1976, Als-Nielsen1976, Mook1981}.   Instead of using the normalized magnetization $M(T)/M(0)$, which is assumed to be aligned with the small external magnetic field $\bm h= \{ 0, \ 0, \ h_z\}$, we adopt a more convenient from the theoretical viewpoint approach to use the normalized  $\hat {\bm z}$-projection of the mean spin $\overline S_z(T)/ S= M(T)/M(0)$. The mean spin $\overline {S}_z$  is defined as
\begin{equation}\label{meanS} \overline {S}_z=\frac{1}{N_{\rm lat}   }\Big\langle \sum_j S^z_j \Big \rangle  \ .
\end{equation}
Here  $S^z_j$  is the local spin projection
on the external magnetic field, $j$ runs over magnetic ions lattice sites per unit volume and $N_{\rm lat}$ is their number.

 In Figure \ref{f:2}(a), we plot the experimental results for the normalized mean spin $\overline S_z(T)/S$   for EuO (main panel, a solid black line with circles) and EuS (inset, a solid blue line with squares). The experimental Curie temperatures $T_{_{\rm C}}^{\rm exp}=69.2\,$K for EuO and $T_{_{\rm C}}^{\rm exp}=16.6\,$K for EuS are clearly visible in the figure. Note that the properties of EuO and EuS are sensitive to stoichiometry and the presence of defects (see, for example, \cite{Borukhovich1976,Pan2012}). In this work, we do not address these technological issues. We refer to the experiments in which the material properties are well controlled. 

The experimental dependencies of $\overline S_z/S$ on the normalized temperature $T/T_{\rm C}^{\rm exp}$ for both materials practically coincide, as is observed in Fig. \ref{f:2}(b). This allows us to focus on the theory-experiment comparison mainly for one material. For concreteness, we choose EuO. The relevant parameters of EuO and EuS are listed in Tab.\,\ref{t:1}. 

The rest of the paper is organized as follows. 

\begin{table} 
    \centering
    \begin{tabular}{c |   c|  c c | c  c   c  | c c  } 
    \hline
     &  $ M_0$  & $ J_1$, & $J_2$  & $T\Sb C\sp{exp}$ & $T\Sb C\sp{th}$ & $\widetilde T\Sb C\sp{th}$   & $\beta_{\rm exp}$ & $\beta_*$ \\
  &   Oe  & K  & K & K & K & K &  & \\ \hline
EuO    &  1920 & $1.25 $ & $0.25 $  &   69.2 & 86.6 & $66.5$  &    $0.36\pm 0.01 $ & $0.34\pm 0.02 $\\
EuS    &  1115 & 0.44 & -0.2 & 16.6 & 21.4 & $13.7$ &   $0.36\pm 0.01 $ & $0.34\pm 0.02 $  \\
   \hline
           \end{tabular}
    \caption{\label{t:1} Important parameters of EuO and EuS: the magnetization at zero temperature  $M_0$;  nearest neighbors $J_ 1$  and next-nearest neighbors $J_2$ exchange integrals;  the experimental value 
 $T\Sb C\sp{exp}$ of the Curie temperature;   the ``theoretical" $T\Sb C\sp{th}$ Curie temperature in the quantum Weiss-Heisenberg  MFA, \eqref{Tc2A};  Curie temperature $\widetilde T\Sb C\sp{th}$, Eq.\eqref{tildeT_c} in the spin-improved MFA. Experimental \cite{Als-Nielsen1976}, $\beta_{\rm exp}$ and theoretical $\beta_*$ values of the apparent scaling index $\beta$ that governs the temperature dependence of the magnetization $M(T)\propto (T_{_ {\rm C}}-T)^{\beta}$ below $T_{_ {\rm C}}$.}
\end{table} 

Section \ref{s:History} is devoted to the historical and physical background of the problem.
 In Section \ref{ss:MFT}, we provide a brief overview of the achievements and fundamental problems of the celebrated classical Weiss mean-field approximation (MFA)\,\cite{Weiss1907}   and recall in Sec. \ref{ss:Q-Weiss} their resolution by Heisenberg, who introduced the exchange interaction of quantum-mechanical origin into the original Weiss MFA \cite{Heisenberg1928}.
 
The quantum version of the Weiss-Heisenberg (WH)  MFA, as represented by \eqref{QWEa}, provides a simple yet reasonable description of the temperature dependence of magnetization (or the mean value of the spin projection on the external magnetic field $\overline S_z$). 
 A numerical solution of Eq.\,\eqref{QWEa} for EuO, plotted in the main panel of Fig. \ref{f:2}(a) by a dotted blue line labeled (1),
gives the WH-MFA value of the Curie temperature (where $\overline S_z=0$)  $T_{_{\rm C}}^{^{\rm WH}}\approx 86.6\,$K for EuO (about 20 \% larger than its experimental value $ T_{_{\rm C}}^{\rm exp}\approx 69.2\,$ K).

The numerical solution for EuS, shown in the inset in Fig. \ref{f:2}(a), yields $T_{_{\rm C}}^{^{\rm WH}} \approx 21.4\,$K (to be compared with  $T_{_{\rm C}}^{\rm exp} \approx 16.6\,$K).

  However, some problems with the WH-MFA still remain.
For example,
  in the low-temperature limit, when $S-\overline S_z\ll S$, Eq.\,\eqref{QWEa} predicts exponential decay of $ \overline S_z $ with $T$, while a well established spin-wave theory gives $S- \overline S_z \propto T^{3/2}$, see for example\,\cite{Baryahtar1983}. 
  
  Vaks, Larkin, and Pikin solved this problem \cite{Vaks1968a,  Vaks1968b} using a developed diagrammatic technique (DT) for ferromagnetics in thermodynamic equilibrium, as briefly outlined in Appendix \,\ref{ss:DT}. 
However, their approach resulted in an unphysical behavior of $\overline S_z(T)$ near $T_{_{\rm C}}$, where the calculated corrections to $\overline S_z$ become infinite.

 To resolve this issue and obtain a regularized description of $\overline S_z(T)$ across the entire range of $T$ from $T=0$ to $T=T_{_{\rm C}}$, we develop in  Appendix \,\ref{s:DT} the DT for spin operators, based on the functional representation of the generating functional ${\cal Z}(\bm h)$, introduced and analyzed in Sec.\,\ref{ss:FR}. 
The first-order correction in the inverse coordination number $1/Z$ to the WH-MFA $\delta \overline S_z(T)\simeq S/Z$ may be found in the one-loop approximation for the effective potential formulated in Sec.\,\ref{ss:1-loop}.  The resulting  Eq.\,\eqref{I1.22A} is presented in Sec.\ref{ss:fin}.
The numerical solution of these equations for EuO is shown in Fig.\,\ref{f:2}(a) by the solid blue line labeled (2).  It decays much faster than the blue dotted line for the WH-MFA with its exponential decay from the value $S$ (not discernible for $T\lesssim 15\,$K). More detailed analysis [not shown in Fig.\,\ref{f:2}(a)] indicates that  the difference  $S-\overline S_z$  is proportional to $T^{3/2}$,
as expected from the low temperature suppression of $\overline S_z$ by spin waves; see, e.g.\,\cite{Baryahtar1983}.

In addition, we observe that the numerical solution of Eq. \eqref{I1.22A}  is in good quantitative agreement with the experiment conducted in EuO (solid black line with circles) up to about 65 K. At this temperature,  $\overline S_z(T)$ decreases twice, reaching S/2.

However, for larger $T$  [$T\gtrsim 65\,$K in EuO]   Eqs.\,\eqref{I1.22A}  give a slower decrease of $\overline S_z(T)$ than in the experiment, with the same  $T_{_{\rm C}}^{^{\rm WH}}$
 ($\approx 86.6$\,K  for EuO) as in the WH-MFA. The reason for this inconsistency is explained in Sec.\,\ref{ss:BL-DT}, which provides a brief overview of the Belinicher-L'vov (BL) DT \cite{Belinicher1984}. Equations \eqref{I1.22A} with $1/Z$-corrections  do not adequately consider the impact of the spin wave on the average spin projection $\overline S_z(T)$ for temperatures close to      $T_{_{\rm C}}$ (specifically, between $0.8 T_{_{\rm C}} $ and $T_{_{\rm C}}$). The key advantage of the BL DT is that it takes into account, order-by-order, the kinematic relationship \eqref{KI} between the spin correlations, which relates transverse spin correlators describing propagating spin waves and longitudinal correlators [including $\overline S_z(T)$].  As shown in Sect.\ref{ss:SW}, this allows us to account for the effect of spin waves even in the zero-order approximation, i.e., in the MFA. 
 
 The numerical solution of the resulting spin-wave-improved WH-MFA \eqref{1.20},
 shown in Fig.\ref{f:2}(a) for EuO by a dashed green line labeled (3) demonstrates much better agreement with the experiment than all previous approaches. In particular, it includes low-temperature spin-wave corrections, proportional to $T^{3/2}$, and it coincides in this respect with $1/Z$-corrected MFA \eqref{I1.22A}. In addition, it decreases much faster than the solution of 
Eq. \eqref{I1.22A} with $T$ increasing toward $T=T_{_{\rm C}}$, in agreement with the experimental behavior of  $\overline S_z(T)$. As a result, it reaches zero at $\widetilde T_{_{\rm C}}^{\rm th}$ (66.5\,K for EuO), which is essentially closer to  $T_{_{\rm C}}^{\rm exp}$
(69.2\,K for EuO) than the previous result (86.6\,K for EuO). The spin waves are highly excited in the vicinity of  $T=T_{_{\rm C}}$, suppressing $\overline S_z(T)$ according to the kinematic relationship \eqref{KI}. Therefore, the mean-field value [proportional to  $\overline S_z(T)$] is smaller, and consequently, the value of $T_{_{\rm C}}$ decreases.
 
Further improvement of our results for $\overline S_z(T)$ is given in Sect.\,\ref{sss:comp}, where we present the $1/Z$-corrected  spin-wave-improved MFA, summarized in  Eq.\,\eqref{eq5}.  The numerical solutions of these equations are shown by a solid green line labeled (4) in the main panel of Fig.\ref{f:2}(a) for EuO and in the inset for EuS.  The calculated Curie temperatures remain identical to those in the uncorrected spin-wave-improved scenarios, as follows from   Eq. \eqref{res3}. However, $1/Z$-corrections significantly improve the behavior of $\overline S_z(T)$ at intermediate temperatures, bringing it much closer to experimental results. This improvement for EuO is evident in Fig.\,\ref{f:2}(a) by comparing the dashed (3) and solid  (4) green lines.
 
Small discrepancies between the calculated and observed Curie temperatures may stem from the approximate nature of the theory, uncertainty of the exchange integrals, or limited accuracy of the Curie temperature measurements. Leaving these differences aside, we plotted in Fig.\ref{f:2}(b) the temperature dependencies of $\overline S_z(T)$ versus normalized temperature $T/T_{_{\rm C}}$, using $T_{_{\rm C}}= T_{_{\rm C}}^{\rm exp}$ for the experimental curves and their own values of $T_{_{\rm C}}$ for the numerical curves. 
We observe excellent quantitative agreement between the theoretical dependence $\overline S_z (T/T_{_{\rm C}})$, shown by the solid green line, and the experimental results, represented by the solid black line with circles for EuO and the blue line with squares for EuS. All three lines coincide in the entire range of temperatures from $T=0$ to  $T\approx T_{_{\rm C}}$. 

It is important to emphasize that the power-law scaling of the magnetization is commonly measured in the reduced temperature range $(1-T/T\Sb  C)\lesssim 0.1$ corresponding to about a half of the reduced magnetization range $M/M(0)\lesssim0.5$. This definition does not reflect the physical origins of $M(T)$ behavior. We distinguish between the intermediate precritical range of temperatures $ 0.7\,  T_{_{\rm C}}  \lesssim T <  T_{_{\rm C}} $, where $\overline S_z/S$ decreases from about $0.7$ to about $0.1$ and the critical range with $\overline S_z/S\lesssim 1/Z\approx 0.1$. In the precritical range, the experimental and theoretical curves of $\overline S_z(T) $  closely follow the power-law-like behavior $\overline S_z(T)\propto (T_{_{\rm C}}-T)^{\beta_*}$ (plotted in Fig.\ref{f:2} as a solid red line).  The apparent precritical anomalous scaling index is estimated to be $\beta_* \approx 0.34 \pm 0.01$, which is in good agreement with the experimental value $\beta_{\rm exp} = 0.36 \pm 0.015$ reported in \cite{Als-Nielsen1976}. Moreover,  it agrees well with the critical scaling index $\beta\sb{cr}\approx 0.365$ derived from the renormalization group theory for the 3D Heisenberg model\,\cite{Guillou1977}.

 We summarize our results in Sect.\,\ref{s:sum}.


 \section{\label{s:History} Historical and physical background}

\subsection{\label{ss:MFT} Classical  Weiss  mean-field approximation} 
The theoretical description of ferromagnetism has a long history, starting with the celebrated  Weiss's mean-field approximation first published in 1907 \cite{Weiss1907}.
 Shortly before this, Langevin developed his theory of paramagnetism, based on the fundamental idea that the orientation of a
molecular dipole of moment $\mu$ in a field $H$ is governed by the Boltzmann distribution law. 
If so, the magnetic momentum per unit volume [the magnetization $M(T)$] is given by the expression
\begin{equation}\label{Langevin}
M=M_0 L\Big(\frac{\mu H} T\Big)\,, \ M_0=N_{\rm lat} \mu\,,  \  L(x) =\coth x- \frac 1x\,,
\end{equation}
where $M_0$ is the magnetization  at $T\to 0$, $L(x)$ is the  Langevin function\,\cite{Landau1970}, $\mu=S \mu\Sb B$, $S$ is the spin of the magnetic ion and   $\mu\Sb{B}=\hbar e/2 mc$   is the Bohr magneton, where $\hbar$ is the reduced Planck constant, $e$ and $m$ are the electron charge and mass, and $c$ is the speed of light.

 The basic idea of the Weiss MFA is that the effective field acting on an elementary magnet in a ferromagnetic medium is not the applied field $H$, but rather $H+g M(T)$, where $M(T)$ is the magnetization at a given temperature and $g$ is some temperature independent factor.  The term $g M(T) $ is called the ``self-consistent molecular field" and is clearly a manifestation of some cooperative phenomenon. 

With this modification, Eq.\,\eqref{Langevin} becomes 
\begin{subequations}\label{Langevin2}
\begin{equation}\label{Langevin2A}
M=  M_0 L [\mu (H+g M) / T]\,.
\end{equation}
 For small $T$,  Eq.\,\eqref{Langevin2A} describes magnetization saturation at the level $M_0$. For very large $T$, the argument $x$ of the Langevin function becomes small and $L(x)$ can be approximated as
 \begin{equation} \label{Langevin2B}
 L(x)  = \frac x3 - \frac {x^3}{45}+ \dots\,,\quad \mbox{for} \ x\ll 1\ .
 \end{equation}
 Accounting only for the first term in  expansion \eqref{Langevin2B}, we reduce Eq.\,\eqref{Langevin2A}  to  
\begin{equation}\label{Langevin2C}
\chi =  M/H\approx\frac{ M_0\mu }{[  3 (T-T_{_{\rm C}})]}\,, \quad  T_{_{\rm C}}  = g\, \mu M_0  /3\ .
\end{equation}\end{subequations}
 Here $\chi$ denotes the susceptibility $M/H$, which formally diverges at some critical temperature  $T_{_{\rm C}}$ known as the ``Curie-Weiss temperature". 
 
\begin{figure}[t]  
  \includegraphics[width=1.\columnwidth]{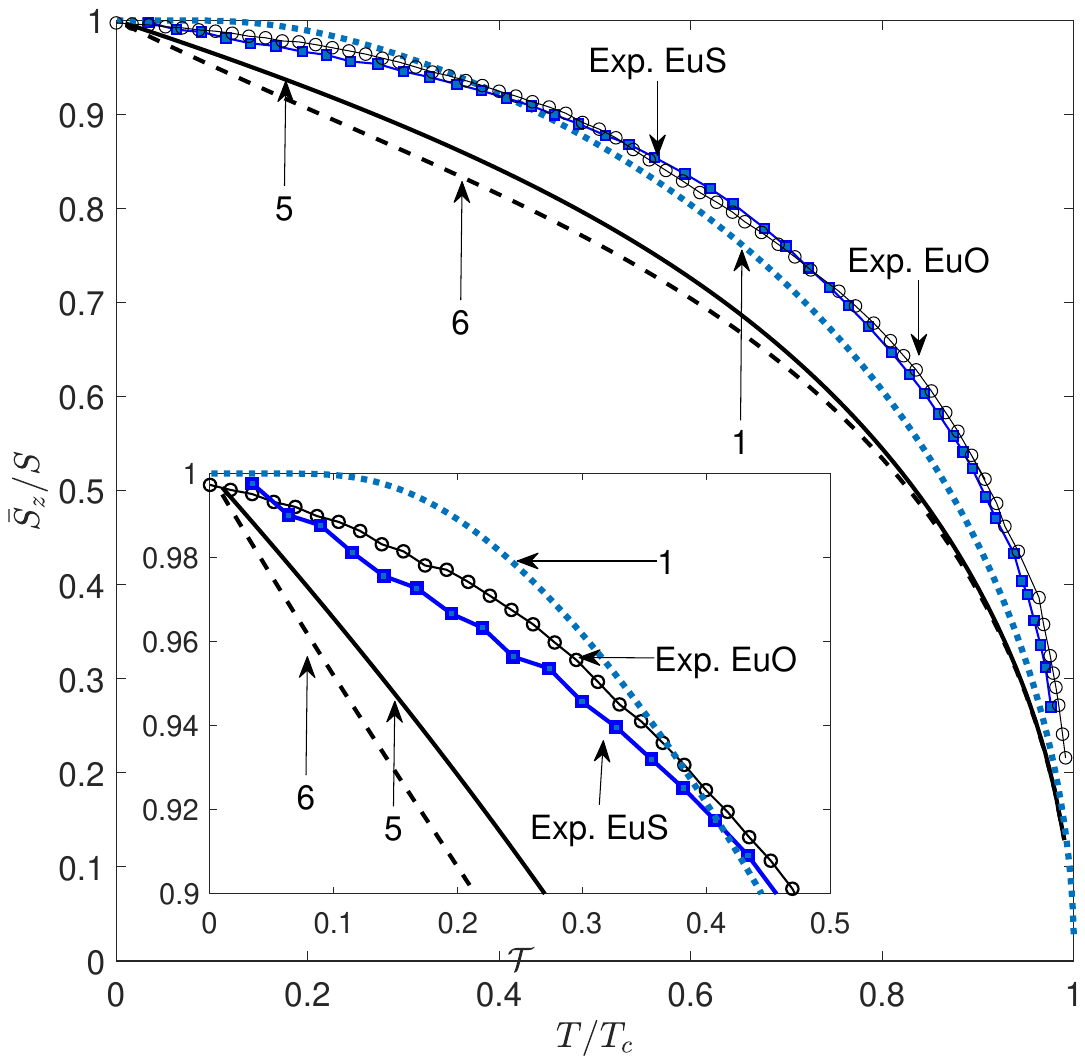}  
 \caption{\label{f:3} Comparison between magnetization temperature dependencies  $\overline S_z/S $  vs. normalized temperature $T/T\Sb{C}$  obtained from:\\
(1) numerical solution of the WH-MFA \eqref{QWEa} - blue dotted line, the same as line (1) in Fig.\ref{f:2};\\
 (5) classical Weiss MFA, numerical solution of Eq.\,\eqref{WeissB} --solid black line;\\
(6) interpolation formula \,\eqref{WeissA} -- dashed  black line;\\
 Experiments in EuO (solid black line with circles) and EuS (solid blue line with squares) are shown by the same line types as in Fig.\ref{f:2}.
  }
\end{figure}

 For $H=0$ and $T$ slightly below $T_{_{\rm C}}$,  Eqs.\,\eqref{Langevin2A} and 
 \eqref{Langevin2B} give 
 \begin{subequations}\label{Weiss}
 \begin{equation} \label{WeissA}
 ~ \hskip - .2 cm  M(T) = M_0  \sqrt{\Big(1+\frac {2{ T}^2}{3 T\Sb C} \Big)\Big (1- \frac{ T}{T \Sb C}\Big )}    \  .
\end{equation}
The interpolation formula\,\eqref{WeissA}, plotted in Fig.\,\ref{f:3} as a  dashed  black line labeled $(6)$, is exact in the limit  $T\to T\Sb C$  and normalized such that $ \overline S_z(0)=S$. It is very close to the numerical solution  of Eq.\eqref{Langevin2A} for $H=0$ which takes the form 
\begin{equation} \label{WeissB}
 {  M(T)}  =M_0  L \Big[\frac{3  M} {M_0}\,     \Big ( \frac{  T}{T\Sb C}\Big ) ^{2/3} \Big ] \,,
\end{equation}\end{subequations}
see solid  black line labeled $(5)$ in Fig.\,\ref{f:3}. 

Undoubtedly, Eqs. \eqref{Langevin2} and 
\eqref{Weiss} represent the most significant result of Weiss's theory. They predict the critical temperature $T_{_{\rm {C}}}$.  As this temperature is reached from below,  ${ M(T)} $ gradually decreases to zero. Beyond this temperature, $ M(T)$  vanishes, consistent with observations from numerous experiments.

However,  in 1907, when   Weiss published his paper\,\cite{Weiss1907}, there was a problem with the obtained values of $T_{_{\rm C}}$, Eq.\,\eqref{Langevin2C}. At that time, the only known interaction between magnetic moments was the classical dipole-dipole interaction, leading to the so-called demagnetization magnetic field, which depends on the shape of the sample. For example, for the orthogonally magnetized film $g=4\pi$, and for the spherical sample $g= 4\pi/3$. Taking for concreteness $g= 4\pi/3$, and actual  EuO values  $M_0\approx 1920 \,$Oe and $\mu = 7  \mu\Sb{B}$, Eq.\,\eqref{Langevin2C} gives $T_{_{\rm C}}\approx 3.77\,$K which is far below its experimental value $T_{_{\rm C}}^{\rm exp}\approx 70\,$K.

The situation is even worse for Yttrium Iron Garnet (YIG),  the ferrimagnetic widely used in fundamental studies\,\cite{Cherepanov1993} and applications\,\cite{Mohseni2022}. With $M_0\approx 155\,$Oe,  $\mu=5\mu_{_{\rm B}}$
 $g= 4\pi/3$ (for the sphere), Eq.\,\eqref{Langevin2C} gives  $T_{_{\rm C}}\approx 0.22\,$K which has nothing in common with the experimental value $T_{_{\rm C}}^{\rm exp}\approx 560\,$K. However, Weiss was courageous enough to publish his article despite the significant discrepancy between the predicted and experimental values of $T_{_{\rm C}}$.
As Van Vleck wrote\,\cite{Vleck1945}, Weiss' approach is ``qualitatively right but quantitatively wrong and is based half on theory and half on the genius at empirical guessing."


 \subsection{\label{ss:Q-Weiss}Exchange interaction and quantum Weiss-Heisenberg  theory}
 The way out of this discrepancy was found 20 years later in the framework of newly emergent quantum mechanics. In 1926, Heisenberg explained\ that, in addition to the magnetic dipole-dipole coupling, a much stronger coupling of Fermi particles-- electrons-- of electrostatic Coulomb nature exists\,\cite{Heisenberg1926}. In quantum mechanics, the wave function $\Psi$ of two identical electrons must be antisymmetric. Therefore, when the spins are parallel, the coordinate part of the $\Psi$ function will also be antisymmetric, while for antiparallel spins, it will be symmetric. This difference in symmetry of the coordinate function $\Psi$ leads to distinct spatial distribution of the two electrons, resulting in a variation in their Coulomb energy, termed by Heisenberg the exchange energy\,\cite{Heisenberg1926,Landau1970}. He proposed a straightforward form of exchange energy $E_{ij} ^{\rm ex}$ between two localized spins $\B S_i$ and $\B S_j$ at lattice points $\B R_i$ and $\B R_j$: 
\begin{subequations}\label{Eex}\begin{equation}
\label{EexA}
E_{ij} ^{\rm ex}  = -  J_{ij}  \B S_i \cdot \B S_j \ .
\end{equation}
 Here $ J_{ij}$ is the so-called exchange integral. Total exchange energy in the lattice $E _{\rm ex}$ reads
\begin{equation}\label{EexB}
E  _{\rm ex}  = -  \frac12\sum_{i,j}J_{ij}  \B S_i \cdot \B S_j \ .
\end{equation}\end{subequations}
Factor $1/2$ accounts for each particular contribution in\,\eqref{EexA}
 appearing in \eqref{EexB} twice. 
 
 The magnetic moment of Eu$^{+2}$ originates from very localized $4f$
electrons with total spin $S=  7/2$ and magnetic moment $\mu = 2 S \mu_{_{\rm B}}$.
In the quantum era, to compute $\langle M(T)\rangle $ we have to account for a discrete series of spin orientations
rather than a continuous distribution, as in
the classical Langevin theory.  With this modification, we have replaced the Langevin function $L(x)$ in \eqref{Langevin2A} by  so-called Brillouin function 
\begin{subequations}\label{Bri}
\begin{align}\begin{split} \label{BriA}
\hskip - .2 cm {\cal B}_{_S}(x)=& \frac{2 S+1}{2S}\coth \!\Big  ( \frac{2 S+1}{2S} x\! \Big )- \frac 1{2S}
\coth\Big  ( \frac{x}{2S}\Big )\\
=& \frac{(1+S)x}{3 S }  + [1- (1+2 S)^4]\frac{  x^3}{720 S^4}+ \dots\ . \    
\end{split}\end{align} 
Now \eqref{Langevin2A} is amended as follows
 \begin{align}   \label{BriB}
M(T)= & M_0 B_{_S} [ \mu (H+  g_{\rm ex} M) / T]\,, \quad \mu= 2S \mu_{_{\rm B}} \,,\\ \nonumber 
g_{\rm ex}= & \frac{J_0}{2 N\sb {lat} \mu _{_{\rm B}}^2}\,, \  J_0=\sum _{j} J_{ij}= Z_1 J_1+ Z_2 J_2 + \dots \ . \end{align} \end{subequations}

Here, the parameter $ g_{\rm ex}$ originates from the exchange interaction\,\eqref{EexA}. $J_0$ is the zero Fourier component of the exchange integral; $J_1$ is the exchange integral between the nearest-neighbor (nn) sites,
$Z_1\equiv Z$ is the nn coordination number (the number of nn pairs); $J_2$ is the next-nearest neighbor (nnn) integral, and $Z_2$ is the nnn coordination number, etc.

The exchange interaction in EuO and EuS occurs indirectly via more extended $5d$ wave functions \cite{Burg2012}. 
Only two types of exchange interactions are important: $J_1$ and $J_2$; for their values, see Tab.\,\ref{t:1}.
The rest of the interactions can be peacefully neglected. Therefore, in \eqref{BriB}  for $J_0$, it is enough to account for only two terms. In FCC crystals, like EuO and EuS,
$Z_1=12$ and $Z_2=6$, see Fig.\,\ref{f:1}.  The lattice separations $\bm R_j=\bm r_{0,j}$ for the $ 12$ nearest neighbor sites  and for the $ 6$ next-nearest  neighbors sites   are:
\begin{align}\begin{split}\label{Rj}
\bm R_1\,,  ... \bm R_{12}  =& \{  \pm  \frac a2, \pm \frac  a2, 0\}\,, \{  \pm  \frac a2, 0,\pm \frac  a2   \}\,, \\
&  \hspace{2.25cm}     \{ 0,  \pm  \frac a2, \pm \frac  a2  \}\, ; \\
\bm R_{13}\,,  ... \bm R_{18}  =& \{  \pm a, 0, 0\}\,, \ \{ 0 \pm  a, 0    \}\,, \{ 0,  0,\pm  a    \}\, .
\end{split}\end{align}
Here $a$ is the size of the full cube in Fig.\,\ref{f:1} consisting of 4 elementary cells of volume $v=a^3/4$ each.

 It is convenient to rewrite \eqref{BriB} in terms of $ \overline S (T)\equiv \langle S  \rangle_{{_T}}$, introducing the so-called ``normalized Brillouin function 
 \begin{align} \begin{split}\label{bS}
 b_{_{S}}(x)=&S {\cal B}_{_{S}}(S x)\\ =  &
   \Big( S+\frac12 \Big) \coth  \Big  (  S+\frac12 \Big)x  - \frac 12
\coth\Big  ( \frac x2 \Big ) \ .
  \end{split}\end{align}
 For small $x$
 \begin{align}\begin{split} \label{b-exp}
   b_s(x)=& \frac{S(S+1)x}{3}  -  \frac{S x^3}{90}B+\dots   \\
B=& 1+ 3S + 4 S^2 + 2 S^3\ .
\end{split}   \end{align} 
 
 For $H=0$, we come to the quantum WH equation 
   \begin{equation}\label{QWEa}
 \overline S (T)=b_{_{\rm S}}\Big[ \frac{ \overline S (T) J_0 }T \Big ]\ .   \end{equation} 
 Using expansion \eqref{BriA} and Eq.\eqref{BriB} for $g_{\rm ex}$, we find a new
equation for the {WH temperature $T_{_ {\rm C}}$ similar to \eqref{Langevin2C}, but now accounting for the exchange interaction \eqref{EexB}:

\begin{subequations}
\begin{equation}\label{Tc2A}
 T_{_{\rm C}}^{\rm th}= \frac{2 S(S+1) N \sb {lat} \mu_{_{\rm B}}^2 g_{\rm ex} }{3 S}=\dfrac{ S (S+1)}3 J_0\ . 
\end{equation} 
Temperature dependence of the magnetization (or mean spin $\overline S$) near $T\Sb C$ is also similar to \eqref{WeissA}, but with a different prefactor
\begin{align} \label{Tc2B}
\overline S=& \sqrt {\frac{T\Sb C}{T}-1} \sqrt \frac {90}{S B} \Big ( \frac  T {J_0}\Big )^{3/2}\ .\end{align}
For $T\approx T\Sb C$ this relation may be approximated as
\begin{align} \label{Tc2C}
\overline S
\approx & \sqrt {T\Sb C -T}\sqrt \frac {90\, }{SB }\frac {T\Sb C } {J_0^{3/2}}\propto \sqrt {T\Sb C -T}\,,
\end{align}
corresponding to the scaling behavior of the order parameter  $\overline S\propto (T\Sb C -T)^{\beta\Sb{MFA}} $ with the ``normal" scaling exponent $\beta\Sb{MFA}= 1/2$  predicted by the Landau theory of the second-order phase transitions\,\cite{Landau1970} in general and MFA in particular. 
\end{subequations}

Taking exchange integrals from Tab.\,\ref{t:1} and using \eqref{BriB}, we obtain $J_0\approx 16.5\,$K for EuO and $J_0\approx 4.1\,$K for EuS. These further give $T_{_{\rm C}}^{\rm th}\approx 86.6$ for EuO and  $T_{_{\rm C}}^{\rm th}\approx 21.4$ for EuS which are not far from corresponding experimental values $T_{_{\rm C}}^{\rm exp}\approx 69.2 \,$K 
for EuO and $T_{_{\rm C}}^{\rm exp}\approx 16.6 \,$K 
for EuS.

The solution of  Eq. \eqref{QWEa}} in the WH-MFA across the whole temperature range for EuO is shown in the main panel and for EuS in the inset of Fig.\,\ref{f:2}(a)  by a blue dotted line labeled $(1)$. However, the agreement with the experiment is only qualitative.
 We conclude that the WH-MFA with the quantum-mechanical Heisenberg exchange interaction, Eq. \eqref{QWEa}, can serve as a leading order approximation for the study of the thermodynamic properties of ferromagnetics. 

Further efforts to improve the WH-MFA took into account larger clusters. In the paramagnetic phase (above $T\Sb C$) their equilibrium dynamics were rigorously studied by Chertkov and Kolokolov\,\cite{Chertkov1994,Chertkov1995}. 
Below $T\Sb C$, larger clusters were studied, e.g., by Chamberlin\,\cite{Chamberlin2000}. However, due to divergence of the correlation length in the proximity of $T_{_{\rm C}}$,  very large clusters are required to achieve the desired precision, resulting in minimal or no computational benefits compared to the evaluation of the whole system.

It is crucial to recognize that the methods mentioned above are unsystematic, making it difficult to control the nature of the assumptions and calculate corrections regularly. This issue can be resolved by using perturbation theory with
graphical notation for the terms, known as the diagrammatic technique.

 The details of the DT usage are too complex for the general reader, as it is geared towards experts in theoretical physics. Therefore, we placed our derivation of required corrections to the MFA in Appendix\,\ref{s:DT}, where the interested reader will find all the technical details of the theory.
 The main physical results of Appendix \ref{s:DT}  are collected and thoroughly discussed in Section \ref{s:beyond},  where we describe the consistent step-by-step improvements of the WH-MFA, culminating in an accurate quantitative description of the temperature dependence of magnetization throughout the entire temperature range from  $T = 0$  to  $T \lesssim  T\Sb C$. These findings are in excellent agreement with experimental results for EuO and EuS.

\section{\label{s:beyond} Beyond the quantum Weiss-Heisenberg MFA}
 
In this Section, we describe systematic, step-by-step corrections to WH-MFA, analyze them in various limiting cases, and explain the physical mechanisms behind the improved description of $M(T)$.

Note that both Weiss and Weiss-Heisenberg MFA replace the actual, time-dependent effective magnetic field $H_i(t)$   acting on some spin $S_i$ with its mean value $H= \overline{ H_i(t)}$, completely neglecting the fluctuations of the surrounding spins $S_j(t)$. The corrections $\delta M(T)$ due to fluctuations are relatively small with a smallness parameter $1/Z$, where $Z=12$ is the coordination number in EuO and EuS.  The initial step to improve WH-MFA is described in Sec\,\ref{ss:fin}, where fluctuations are considered in the first order of perturbation theory with respect to $1/Z$. This allows us to correctly describe the power-like decay of $M(T)$ for $T \ll T\Sb C$    
caused by spin waves instead of the incorrect exponential decay of $M(T)$ in the WH-MFA.

Nevertheless, the behavior of $M(T)$ near $T\Sb C$ is still not corrected sufficiently and gives the same $T\Sb C$ as the initial WH-MFA. This problem is addressed in the subsequent sections: \ref{ss:BL-DT} and \ref{ss:SW}. The effect of long-propagating spin waves on the fluctuations of   $H_i(t)$ is described more accurately by considering the exact kinematic identities \eqref{KI}, which connect all projections of the spin operator  $\widehat{\bm S}$. This approach improved the behavior of  $M(T)$ not only in the low-temperature range but also near the temperature  $T\Sb C$, including the value of $T\Sb C$ itself.   

In the final Sec.\ref{ss:IIIF}, we combine two types of corrections to get the accurate quantitative description of $M(T)$ in the entire temperature range from $T=0$ to $T\Sb C$.
 \subsection{\label{ss:fin} $\dfrac 1 Z$-corrections to the Weiss-Heisenberg MFA}
  The perturbation theory in $1/Z$  works well if the corrections $\delta M(T)$ due to fluctuations of the magnetic fields are small compared to the mean value $M(T)$.  
 Unfortunately, the fluctuations of the effective magnetic field $H_i(t)$ on a given spin $S_i$ are relatively small with respect to their mean value $H$ only when $T\ll T\Sb C$. As $T$ approaches $T\Sb{C}$, $H$ vanishes.  In this case, the $H_i(t)$ fluctuations become larger compared to $H$, even in the first order in $1/Z$. This is a common problem in the theory of second-order phase transitions. 
 First attempts in this direction, even using the diagrammatic perturbation approaches, faced serious problems:  the VLP perturbation approach leads to infinite values of $1/Z$ corrections when $T\to T\Sb C$   (for more details, see Appendix \ref{ss:DT}). As is  
 elaborated in Appendix\,\ref{ss:FR}, we use a more sophisticated version of DT based on a generation functional. In our version of DT, $1/Z$ corrections are introduced in a much more compact form of one-loop effective potential, described in Appendix\,\ref{ss:1-loop}. The resulting equations with the first-order in $1/Z$  corrections  are as follows: 
\begin{align} \begin{split} \label{I1.22A}  
\overline S=&   A_0  +  A_1  +  A_{2}  +    A_3\,, \\  
   A_0  =&  b_s(\beta J_0 \overline S)  \,,  \quad   A_1 =  A_{1a} + A_{1b} \,,  \quad \beta=1/T\,,\\
  A_{1a} =&- \big \langle n_{\bm k }\big \rangle_{\bm k } = - {\cal N}\,,   \\
   A_{1b} =&  \big   \langle n_0 (\beta \overline S J_0)    \big \rangle_{\bm k } = [\exp (J_0\overline S/T)-1]^{-1}\,, \\
    A_2  = &   \beta b_S ^\prime (\beta J_0 \overline S) 
\big \langle J_{\bm k}   n_{\bm k} \big \rangle _{\bm k } \,,  \\   
 A_3= & \frac{ \beta b_S ^{\prime\prime}(\beta J_0 \overline S)}{2 }\Big \langle \frac {\beta J_{\bm   k}\, }{1- \beta J_{\bm k }  b_S'(\beta J_0 \overline S)  } \Big \rangle_{\bm k} \ .
\end{split}\end{align} 
 Here
\begin{equation}\label{Mean}
    \big \langle f_{\bm k }\big \rangle_{\bm k }\equiv  \frac{v}{(2\pi)^3}\int f_{\bm k} \, d^3 k\,, \quad  \big   \langle 1\big \rangle_{\bm k }=1\,, 
\end{equation}
which  can be interpreted as the mean value of some function  $f_{\bm k}$ per magnetic site,
$v$ is the unit cell volume and integration in Eq.\,\eqref{Mean} over $\bm k$ is carried out in the first reduced Brillouin zone for the wave vectors.

In Eq\,\eqref{I1.22A},  \,$b_{_S}^\prime(x)=d b_{_S}/d x$ and  $b_{_S}^{\prime\prime}(x)=d^2 b_{_S}/d x^2$ are the first and the second derivatives of the normalized Brillouin function $b_{_S} (x)$ given by \eqref{bS}, and ${\cal N}$ is the mean value of the magnon numbers $n_{\bm k}$ per magnetic site, 
given  by the Bose-Einstein distribution: 
 \begin{align}  \begin{split}
   \label{III.33} n_{\bm k}= & 
 \Big [\exp    \frac{E_{\bm k}(T)}T -1 \Big ]^{-1}\,,  \\ 
E_{\bm k}(T)=&  \overline  S_z(T) (J_0-J_{\bm k}) \,,\\  
J_{\bm k}=&\sum_j J_{ij}\exp (i \bm k \cdot \bm R_{ij})\ .
\end{split}  \end{align}
Here $ E_k(T)$ is “self-consistent” energy of
spin waves, which can be found in the simplest version
of the Green's function splitting suggested by
Tyablikov\cite{Tyablikov1967}.  To rationalize this result from a physical point of view, at least for small $ak\ll 1$ and final temperatures, note that in these conditions, the main contribution to the decrease in $\overline S_z$ comes from the fast spin waves with $ak\sim 1$. This allows us to average the spin system over fast motions and to consider slow spin waves with $ak\ll 1$, as in the limit $T\to 0$, by replacing $S$  (in our case $S=7/2$) with $\overline S_z(T)$.

Note that Eqs. \eqref{I1.22A}, considered as a straightforward expression for $1/Z$-corrections, are  divergent at
$T\to T\Sb C$. However, if we consider   \eqref{I1.22A} as self-consistent equations for $\overline S$, the divergencies are resolved because this procedure corresponds to a partial summation of the most divergent corrections in all orders in  $1/Z$, similarly to Dyson resummation for the Green's functions in many versions of the diagrammatic perturbation approaches.

For actual calculations in EuO and EuS, we need explicit expressions for $J_0-J_{\bm k}$  in these crystals. From \eqref{III.33} and \eqref{Rj} we find:
\begin{align}\begin{split}\label{Ek}
 \hspace{-.1 cm} J_0-J_{\bm k}=& 4   J_1 \Big[\sin^2 \frac {(k_x+k_y)a}{4}+ \sin^2 \frac {(k_x-k_y)a}{4} \\ & +\sin^2 \frac {(k_x+k_z)a}{4}+ \sin^2 \frac {(k_x-k_z)a}{4} 
 \\ & +  \sin^2 \frac {(k_y+k_z)a}{4}+ \sin^2 \frac {(k_y-k_z)a}{4}  \Big]\\
& +  4 J_2 \Big [\sin^2     \frac{k_x a}2  + \sin^2  \frac{k_y a}2+ \sin^2  \frac{k_z a}2  \Big]\ .
\end{split}\end{align}


\subsubsection{\label{ss:T->0}Low-temperature  spin-wave-dominated regime A}
In the limit $T\to 0$, the  term $A_0$ in Eq.\,\eqref{I1.22A} gives the expected trivial answer $\overline S=S$. Together with the term  $A_{1a}$, it gives a well-known result  (cf. \cite{Baryahtar1983}) for the spin-wave correction of low-temperature mean spin
\begin{equation}\label{Sz2}
\overline S\approx S - {\cal N}(T)\,.
\end{equation}  

 To estimate ${\cal N}(T)$ note that at small $T$,
say for $T<10\,$K in EuO, the integral in  $\langle n_{\bm k}\rangle $ is dominated by small $k$ and the upper limit can be expanded to $\infty$.  In cubic crystals, for small $k$
 \begin{subequations}\label{Ek1}
 \begin{equation}\label{Ek1A}
  E_{\bm k}(T)= E_{\rm ex}(ak)^2\ .
 \end{equation}
 For example, for EuO and EuS  [according to      \eqref{III.33}  and Eqs.\,\eqref{Ek}]
\begin{equation}\label{Ek1B}
  E_{\rm ex} = \overline  S_z(T) (J_1+J_2)\ .
 \end{equation}
 \end{subequations} 
 This allows us to integrate $\langle n_k \rangle$ over angles in spherical coordinates and, finally, to come to a one-dimensional integral
 $$
  \int\limits_0^\infty \dfrac {x^2 dx}{\exp x^2 -1  }=\dfrac{\sqrt \pi }{4} \zeta (3/2) \approx 1.579.
  $$ 
  
  Then ${\cal N}(T)$ can be estimated as 
 \begin{equation}\label{N3}
{\cal N}(T)\approx \frac{\zeta(3/2)}{32 (\pi)^{3/2}}\Big ( \frac{T}{ E_{\rm ex}} \Big )^{3/2}\ .
 \end{equation}
Using $J_1, J_2$ from Tab.\ref{t:1} in Eq.\,\eqref{Ek1B}, we obtain the values of $E_{\rm ex}\approx 1.50\,$K  for EuO and $E_{\rm ex}\approx 0.22\,$K  for EuS.

In the limit $T\to 0$ the rest of the terms  in \eqref{I1.22A} give exponentially small corrections: \\
-- Term $A_{1b} \approx \exp(-J_0 S/T)  $;\\
-- Term $ A_2 \approx \exp(-J_0 S/T){\cal N} $ where ${\cal N}   \propto T^{3/2} $, see \eqref{N3};\\ 
-- Term $A_3 \approx \dfrac {\langle J_{\bm k} \rangle_{\bm k}}{2T}\exp \Big(-\dfrac {J_0 S}{T}\Big ) $, where $ \langle J_{\bm k} \rangle_{\bm k} $ is defined by Eq.\,\eqref{III.33}.

 Therefore, our result, Eqs.\eqref{I1.22A}, gives an expected and well-known low-temperature behavior (for $T\lesssim 40\,$K in EuO) of $\overline S(T)$, see, for example,  \cite{Baryahtar1983}.
 \subsubsection{\label{sss:T->Tc} Near-$T_{_{\rm C}}$ behavior of $\dfrac 1 Z$-corrected MFA }
    
 In the critical regime C, the spin fluctuations dominate the magnetization behavior, and even the corrected MFA is no longer valid. Nevertheless, its analysis is still instructive, allowing us to find the estimate of $T_{_{\rm C}}$  in this approximation.
    
    To find the behavior of the magnetization in the limit $\overline S\to 0$, we consider the basic equations\,\eqref{I1.22A} with  $n_{\bm k}$   given by Eqs. \eqref{III.33}. Here, we can use the Rayleigh-Jeans distribution 
\begin{equation}\label{def3}
n_{\bm k} = \frac{ T}{  E_{\bm k}} \Rightarrow \frac{T}{\overline S (J_0 -J_{\bm k} )}\,, \quad n_0(x) \Rightarrow \frac{T}{\overline S  J_0 } \ .
\end{equation} 
Using the expansions  of \eqref{b-exp} for $x=J_0  \overline S/T\ll 1$,
  we can simplify equations for $A_j$ accounting for terms of order of $x^{\pm 1}$,  denoted as $A_j^{\pm 1}$:
\begin{align}\begin{split}\label{exp3}
A_0^{(1)}=& \frac{S(S+1)}3\frac{\overline S J_0}T= \overline S \frac {T\Sb C}{T} \,,\\ 
A_1^{(-1)}=&  - \frac{T}{ \overline SJ_0}\Big \langle \frac {J_k}{J_0-J_k}\Big \rangle\,,\quad A_1^{( 1)}=0\,,  \\
A_2^{(-1)}=&  \frac{S(S+1)}{3 \overline S} \Big \langle \frac {J_k}{J_0-J_k} \Big \rangle\,, \\
A_2^{(1)}=&  - \frac{ B J_0^2 \overline S }{T^2} \Big \langle \frac {J_k}{J_0-J_k} \Big \rangle\,, \\
A_3^{(1)}=&  -  \frac{S B}{15} 
\frac{J_0 \overline S}T \Big \langle \frac {J_k}{T -J_k S (S+1)/3}\Big \rangle \\
\approx &\frac {B J_0 \overline S}{5 (S+1)}  \Big \langle \frac {J_k}{J_0 -J_k} \Big \rangle
\end{split}\end{align}
Here, averaging $\langle \dots \rangle$ is understood according to \eqref{Mean}.   

We see that in the limit $\overline S\to 0$ the terms $A_1^{(-1)}$ and $A_2^{(-1)}$ diverge as $1/\overline S$. However, the sum of these two terms
$$A_1^{(-1)}+A_2^{(-1)}= \Big \langle \frac {J_k}{J_0-J_k}\Big \rangle \frac 1 {\overline S} \Big [\frac {S(S+1)}3 - \frac{T}{J_0} \Big ]\propto \frac{T_{_{\rm C}}-T}{\overline S}
$$
i.e. behave as $\overline S$   near   Curie temperature\,\eqref{Tc2A}, $T_{_{\rm C}}= S(S+1)J_0/3$ under assumption that  $\overline S\propto \sqrt{T\Sb C-T}$.

To verify this assumption, consider Eq.\eqref{I1.22A} with approximate values of $A_j$ given by \eqref{exp3}. To simplify the appearance of the resulting \eqref{res3} we multiply each term by $\overline S $, divide by $\Big \langle \dfrac {J_k}{J_0 -J_k}\Big \rangle $ and rearrange to get
\begin{align}\begin{split}\label{res3}
\frac{B J_0\overline S^2}{T(S+1)} \Big [\frac{3}{S}+ \frac 1{5(S+1)} \Big]  = \frac {T_{_C} - T}{J_0}\ .
\end{split}\end{align}  
Here we note that $\overline S^2 -\overline S A_0^{(1)}$
is proportional to $\overline S^2 (T_{_{C}}-T)\propto \overline S^4$, and we neglect it in  \eqref{res3}. Terms   $A_2^{(1)}$ and $A_3^{(1)}$ contribute to the left-hand-side (LHS) of Eq. \eqref{res3}, while  $A_1^{(-1)}$ and $A_2^{(-1)}$ contribute to its  right-hand-side (RHS).

In particular, we observe that as   $T $ approaches the critical temperature  $T_{_{\rm C}}$, the order parameter  $\overline S$   behaves as  $\overline S \propto \sqrt{T_{_{\rm C}} - T}$ . This demonstrates that within the $1/Z$-corrected MFA the scaling behavior of the order parameter, expressed as  $ \overline S \propto (T_{_{\rm C}} - T)^{\beta_{\rm MFA}} $, with the normal scaling exponent  $ \beta_{\rm MFA} = \frac{1}{2}$ . 
Furthermore, the  $1/Z$   corrections to the original quantum  MFA  do not alter the critical temperature  $T_{_{\rm C}} $  as given in the original equation  Eq. \eqref{Tc2A}. These corrections only affect the prefactor, which now differs from that in Eq. \eqref{Tc2B}.


  \subsubsection{\label{ss:Tc}Numerical solution of Eq.\eqref{I1.22A}}
The numerical solution of Eq.\eqref{I1.22A}, shown in Fig.\,\ref{f:2}(a) by the solid blue line, is compared to the solutions of the Weiss-Heisenberg Eq.\eqref{QWEa} (dotted blue line) and the experimental results for EuO. It is observed that the solution of the mean-field equation $\overline S= b_s(\beta J_0 \overline S $)  deviates significantly from the experiment (solid black line with circles) in the entire temperature range. At the same time, the solution of the $1/Z$-corrected Eq.\eqref{I1.22A} (represented by the solid blue line) offers an accurate quantitative description of the experiment in the low-temperature region ($T\lesssim 60\,$K for EuO), dominated by spin-wave contribution.    
However, this solution deviates significantly from the experiment for larger temperatures. 

A way to resolve this problem, demonstrating very accurate qualitative agreement between the developed theory and experiment, is shown below. 

\subsection{\label{ss:BL-DT} Non-equiribrium Belinicher-L'vov DT for spin operator }
The essential progress in this area was made by Belinicher and L'vov (BL), who developed a diagrammatic technique for spin operators \cite{Belinicher1984} using graphical notations similar to those traditionally employed in Feynman's diagrammatic technique. BL extended Keldysh's diagrammatic technique to non-equilibrium Bose systems for the case of spin operators. The key element here was the formulation of Wick's theorem for spin operators, which allows the expression of the mean values of the product of any number of spin operators
$$
\widehat S_\pm =  (\widehat S_x \pm i \widehat S_y)/ \sqrt{2}\quad \mbox{and}   \quad\widehat S_z
$$ through products of only operators $S_z$.
The BL-DT explicitly incorporates the mean values of the spin-wave propagators (expressed via $\widehat S_\pm$) while the longitudinal correlation functions of $\widehat  S_z$ play a role as external parameters characterizing the media in which the spin waves propagate. Calculating longitudinal correlations is a non-trivial task. 
One way to achieve this is to formulate a perturbation approach starting from the basic WH-MFA in which there is no mention of spin waves. With this starting point, one has to account in any order of perturbation approach for the fulfillment of the kinematic identities 
\begin{equation}\label{KI} 
\widehat{ \bm S}^2 = - 2 \widehat S_+ \widehat S_- + \widehat S_z^2 -    \widehat S_z= S(S+1)
\end{equation}
which provides a connection between dynamic operators  $\widehat S_\pm $ and powers of the static
ones, $\widehat S_z$. 

 In BL-DT kinematic identities, \eqref{KI} is used from the start to express correlations $\langle \widehat S_z^n\rangle$ as polynomials of $\widehat S_\pm$. These correlations can be calculated using the longitudinal part of the DT. 
 This approach allows for an efficient calculation method in which even a single simple diagram considered corresponds to the summation of a series of several diagrams in the framework of the Vaks-Larkin-Pikin approach. 
\subsection{\label{ss:SW}Spin-waves-improved Weiss-Heisenberg MFA}
BL showed that accounting for kinematic identities \eqref{KI} in the leading order approximation in $1/Z$ leads to the following modification of MFA:
 \begin{align} \label{1.20}
 \overline S_z(T)=&   b_{_{\rm S}}(  y)\,, \   y =    \ln \Big (1+ \frac{1 }{\cal N}\Big) ,   
\end{align}
where ${\cal N}= \langle n_{\bm k}\rangle_{\bm k}$, the mean value of the occupation numbers,  is the free parameter of the problem. Its value depends on real physical conditions: in the case of strong external pumping, it can be found from the wave kinetic equations.   In thermodynamic equilibrium, $n_{\bm k}$ is determined by the Bose-Einstein distribution \eqref{III.33}. 

In thermodynamic equilibrium, equations Eq.\,\eqref{1.20} were suggested by  Praveczki\,\cite{Praveczki1961} for $S\leq 3.0$ (recall that in EuO and EuS $S=7/2$),  and they were accounting only for nearest-neighbors interactions, see also \cite{Callen1963,Praveczki1963,Tyablikov1967}.

Eqs.\eqref{1.20} can be rewritten in a form  more closely resembling the original version \eqref{QWEa} of the quantum MFA:
\begin{subequations}\label{newver}
\begin{align} \label{newverA} \overline S_z(T)=&  b_{_{\rm S}} \Big[ \frac{E_{\rm eff}(T)}T \Big]\,,\\ \label{newverB} \begin{split}
\hskip - .15cm \Big [ \exp \frac{E_{\rm eff}(T)}{T} -1 \Big ]^{-1}=&\Big \langle \Big [ \exp \frac 
 {E_{\bm k}(T)}{T}-1\Big ]^{-1}\Big \rangle, \\
 E_{\bm k}(T)=& \overline S_z (J_0 - J_{\bm k})\ .
\end{split}\end{align}\end{subequations}
Here, we replace $\overline S (T) J_0$ in Eq.\,\eqref{QWEa}  with the effective energy on the site $E_{\rm eff}(T)$ defined by  Eq.\,\eqref{III.33}. In a way, $E_{\rm eff}(T)$ can be considered as a sophisticated mean value of the spin waves energy $E_{\bm k}(T)$  over the entire $\bm k$-space at a given $T$.
We consider Eq.\,\eqref{newver} to be physically motivated and more transparent than its original form \eqref{1.20}. 

As is shown below, Eq.\eqref{1.20} [or, equivalently, Eq. \eqref{newver}] is significantly more accurate than the corresponding Eq.\eqref{QWEa} for WH-MFA and even  Eqs.\eqref{I1.22A}  describing WH-MFA with $1/Z$-corrections.

\subsubsection{\label{sss:Tto0}  $ \overline S_z(T)$ in low-$T$ regime A } 
When $T\to 0$,   ${\cal N}$  is very small and $y\gg 1$ in \eqref{1.20}.  In that case
 \begin{equation}\label{Sz1}
 \overline S_z(T)\approx S - \frac{{\cal N}(T)}{1+{\cal N}(T)}\approx S -  {\cal N}(T)\,,
 \end{equation}
 i.e., as anticipated, the  decrease  in $\overline{S}_z(T)$ is precisely governed by the excitation of spin waves\, \cite{Baryahtar1983}. Note that Eq.\,\eqref{Sz1} coincides with Eq.\,\eqref{Sz2} for $ \overline S_z(T)$ in the ``original" mean-field approximation with $1/Z$
 corrections.

 
\subsubsection{\label{sss:lowT}  $ \overline S_z(T)$ in precritical regime B}
Let us show that the temperature dependence $ \overline S_z(T)$ in the spin-wave-improved mean-field approximation, considered here, and in the  ``original" mean-field, $1/Z$ corrected  approximation is very different when  $T$ approaches $T_{_{\rm C}}$, the mean spin $\overline S_z(T)\to 0$
 and the energy $ E_{\bm k}(T)$ in \eqref{newverA} also approaches zero and become much smaller than $T$.  In that case, the Bose-Einstein distribution \eqref{newverB} can be approximated by the Rayleigh-Jeans distribution $T/E_{\bm k}(T)$ and Eq.\eqref{newverB} reduces to 
 \begin{subequations}\label{mean}
 \begin{equation}\label{meanA}
{ E_{\rm eff}= \overline S\widetilde J_0}   \ .  
 \end{equation} 
 Here, we introduce the effective exchange integral, defined as follows 
\begin{equation}\label{def-tildeT}
 \frac 1 {\widetilde J_0}\equiv \<  \dfrac 1 {J_0 - J_{\bm k}}  \>_{\bm k}  .
 \end{equation}     
We observe that Eqs. \eqref{newverA} and \eqref{meanA} coincide with the original mean-field Eq. \eqref{QWEa} after replacing $\widetilde J_0$ with $J_0$. This means that in the spin-wave-improved MFA $\overline S_z$ vanishes at the new value of the Curie temperature
\begin{equation}\label{tildeT_c}
 \widetilde T_{_{\rm C}}= \frac{S(S+1)\widetilde J_0} 3=  \frac{S(S+1) } 3
 \Big \langle \frac 1 {J_0-J_k}\Big\rangle ^{-1}\,, 
 \end{equation}
and in its vicinity it behaves  as a square root of the temperature difference, similarly to \eqref{WeissA}:
\begin{align} \label{S-mean1}
\overline S= C \sqrt {\frac {\widetilde T_{_{\rm C}}-T}{\widetilde T_{_{\rm C}}}}\,, \quad C=\frac {1}{S   }\sqrt {\frac 3 {B (S+1)^3}}\,,
 \end{align}\end{subequations}
where $B$ is defined by  \eqref{b-exp} and $C$ depends only on $S$. As is evident from Eq.\eqref{S-mean1}, the scaling behavior near the Curie temperature within the spin-wave-improved MFA remains normal $\beta\Sb{MFA}=1/2$.  
However, the Curie temperature, denoted as $\widetilde T_{_{\rm C}}< T_{_{\rm C}}$,  is now determined by the effective exchange integral $\widetilde J_0 $, which is smaller than the "bare" exchange integral $J_0$.   The ratio $\widetilde T_{_{\rm C}}/ T_{_{\rm C}}\approx 0.77$ in EuO and
$\widetilde T_{_{\rm C}}/ T_{_{\rm C}}\approx 0.64$ in EuS. The physical reason for this improvement is the suppression of the longitudinal correlations of spins due to the excitation of spin waves in the vicinity of $T_{_{\rm C}}$, as is reflected by kinematic identities \eqref{KI}.
\subsubsection{\label{sss:comp}Numerical solution of Eq.\,\eqref{1.20} and its analysis }
The numerical solution of the spin-wave-improved WH-MFA, \eqref{1.20}  for  $\overline S_z(T)$   in EuO is shown in Fig.\,\ref{f:2}(a) by the dashed green line labeled (3). It is interesting to compare this behavior with the similar curve of  $\overline S_z(T)$ in the $1/Z$ corrected WH-MFA, Eq.\,\eqref{I1.22A}, plotted in Fig.\,\ref{f:2}(a) by the solid blue line labeled (2). We showed analytically that in both approximations, the low $T$ behavior is the same: $\overline S_z \approx S -{\cal N}(T)$. Consequently, both curves practically coincide for $T\lesssim 20\,$K and remain quite close for temperatures up to $T\simeq 50\,$K. However, for larger $T$ these curves deviate significantly: the solid blue line for the $1/Z$-corrected MFA goes to zero at $T\Sb C\approx 86.6\,$K, while the dashed green line for the spin-wave-improved  MFA approaches zero at $\widetilde T\Sb C\approx 66.5\,$K which is much closer to the experimental value $T\Sb C^{\rm exp}$ in EuO.

 We conclude that the spin-wave-improved MFA captures physics better than its 1/Z-corrected counterpart for $T>T_{_{\rm C}}/2$.  We tend to associate this difference with the kinematic identities, \eqref{KI}, that relate the dynamic operators $\widehat S_\pm$ to the static ones $\widehat S_z$ on the same site. In the original WH-MFA, an interaction of a spin on a given site with its real environment with fluctuating spins is approximated by its interaction with the mean values of surrounding spins, producing a time-independent magnetic field $H\sb{eff}= \overline S_z J_0$. In thermodynamic equilibrium, this field causes a non-zero value of $\overline S_z/S$, given by the normalized Brillouin function $b\Sb S$ according to \eqref{QWEa}.
 However, in the presence of intense spin waves, this is not the case: according to kinetic identities when 
 $\langle \widehat S_+\widehat S_-\rangle \ne 0$, the mean spin  $\overline S_z $ cannot reach its maximal value $S$ and vanishes for smaller $T$ i.e.  faster than without accouning for the spin waves.

 
 \subsection{\label{ss:IIIF} $  1/Z$-corrections to the spin-wave-improved mean-field equation and its analysis }
Using the spin-wave-improved mean-field equation, Eq.~\eqref{1.20}, as a  leading order approximation, and finding the required first-order corrections in $\dfrac 1 Z$, we obtain the following version of the self-consistent equation for $\overline S$:
\begin{subequations}\label{eq5}\begin{align} 
\overline S =&  \widetilde A_0 +  \widetilde A_1 +  A_2 + A_3\,, \\
\widetilde A_0=& b_s(y)\,,\quad  y = \ln \Big ( 1+ \frac 1 {\cal N} \Big )\,, \\ \label{eq5C}
\widetilde A_1=& \Big [ 1 -\frac {b^\prime _s (y)}{{\cal N}({\cal N}+1)} \Big]A_1\ .
 \end{align}\end{subequations}
Here $A_1$, $A_2$, $A_3$ and $\cal N$ are given by Eq.\,\eqref{I1.22A}. Equation $\overline S =  \widetilde A_0$ coincides with the spin-wave-improved mean-field equation~\eqref{newver}. As we have shown, it gives for $T\to 0$ 
 the spin-wave correction $\overline S= S -{\cal N}$. The same correction gives the term $A_1$ in Eq.\eqref{I1.22A}.  The prefactor in Eq.\eqref{eq5C} vanishes for $T\to 0$ to prevent double counting of the spin-wave correction.

Note that analysis of near-$T_{_{\rm C}}$ behavior of $\overline S(T)$ in the framework of Eqs.\,\eqref{exp3} for $1/Z$ corrected   WH-MFA shows that  $\overline S(T)=0$ at  $T_{_{C}}= S(S+1)J_0 /3$. This follows from the balance  of $A_1$, $A_2$ and $A_3$
terms   in \eqref{res3}. The analytical equations for these terms in \eqref{eq5},  are the same as in Eqs.\eqref{I1.22A} and \eqref{exp3}. Therefore, it is not surprising that 
$\overline S(T)\to 0$ for $T\to T_{_{\rm C}}$ and not for $T\to \widetilde T_{_{\rm C}}$. 
The reason is that we derived the equations for $A_j$  in the framework of DT developed in Appendices \ref{ss:FR} and \ref{ss:1-loop}, which ignores kinematic identities. The straightforward way to account for them is to 
 derive $A_j$ in the framework of the BL DT. 
  
 Unfortunately, this is quite a cumbersome procedure.  
Instead, we observe that in the range where $\overline S\ll S$, spin waves suppress the effective exchange integrals by a factor of $R\equiv \widetilde J_0/J_0$. Therefore, in all expressions for $A_j$,  we
replace $J_0$ by $\widetilde J_0$ and $J_{\bm k}$ by $R J_{\bm k}$.
    Modified in this way, Eqs.\,\eqref{eq5} for $1/Z$-corrected, spin-wave-improved MFA for $\overline S_z/S(T)$ were solved numerically with Eu0 parameters and shown in Fig.\ref{f:2}(a) by a solid green line.  The accuracy of the numerical solution decreases as $ \overline S$ becomes smaller. We cannot guarantee its reliability when  $S(T) < 0.1 \overline S$.   
  
  Therefore, we did not plot the numerical solution for  $S(T)$  in this $T$-range.
   
 We see further improvement of the result in comparison with the uncorrected spin-wave-improved mean-field equation \eqref{newver}, plotted in Fig.\ref{f:2}(a) by the dashed green line.

When the small discrepancy in the Curie temperature values is  compensated by plotting in Fig.\ref{f:2}(b) $\overline S_z/S$
vs normalized temperature $T/T\Sb {C}$, the theoretical curves of $\overline S_z/S$ in the $1/Z$-corrected spin-wave-improved MFA almost coincide with experimental results for EuO and EuS.  

 Another impressive result is observed in Fig.\,\ref{f:2}. We approximated the numerical solution of  Eqs.\,\eqref{eq5} $\overline S_z/S$  [accurate for $S(T) \gtrsim 0.1\overline S$ ] by a power-law-like $T$-dependence $(1-T/T\Sb {C})^{\beta_*}$ in the temperature range  corresponding to $0.1 \overline S<S(T)< 0.6 \overline S$. This dependence is plotted in Fig.\,\ref{f:2} by the solid red line in the wider $T$-range including $T=\widetilde T\Sb C$, where $S(T)=0$. The value of the apparent precritical anomalous scaling index $\beta_*=0.34 \pm 0.02$ is in excellent agreement with its experimental value $\beta\sb{exp}=0.36\pm 0.01$ in EuO and EuS \cite{Als-Nielsen1976}. 
 Note that the experimental scaling index was measured by fitting $S(T)$ in the $(T\Sb C-T)/T\Sb C\lesssim0.11$ range. 
Therefore, the experimental value $\beta\sb{exp}=0.36\pm 0.01$ corresponds to the precritical regime B.

 It is noteworthy that $\beta_*$ and $\beta\sb{exp}$ are also quite close to the theoretical critical exponent $\beta\sb{cr}\approx 0.365$, deduced within the renormalization group theory for the 3D Heisenberg ferromagnet, particularly suitable in the critical regime C\,\cite{Guillou1977}. 
 While it is not entirely clear why  $\beta_*$ and $\beta\sb{cr}$   are so similar, the extended tail of critical scaling in the precritical regime may be attributed to the particular way we treat the $1/Z$-corrections to  MFA.  Within the spin-wave-improved Weiss-Heisenberg MFA, we consider exact kinematic identities \eqref{KI}, which incorporate complete perturbation series rather than just first in  $1/Z$ terms. Moreover, as was mentioned above, we solve \eqref{eq5} self-consistently.  This process is akin to Dyson's resummation of (presumably) the most divergent terms. It is plausible that the described dual resummation of infinite diagrammatic series captures the key contributions from longitudinal and transversal spin fluctuations, thus extending the scaling from the critical to the precritical regime. Experimental observations matching the extended precritical anomalous scaling lend additional support for this conjecture and the importance of the introduced corrections for the description of basic physical mechanisms defining magnetization.  
\section{\label{s:sum} Summary} 
In this paper, we revisited the theoretical description of spontaneous magnetization in cubic ferromagnetic crystals.
Using a consistent two-step procedure based on the diagrammatic technique, we developed a theory that accounts for intensive and long-propagating spin waves — fluctuations of the transverse spin components.
Our theory resolves the long-standing problem of accurate quantitative description of the temperature dependence of magnetization in cubic ferromagnetic crystals.

We explain the physical reasons for the failure of other approaches and demonstrate the step-by-step improvement in describing the temperature dependence of the spontaneous magnetization emerging in our method. 

Our theoretical approach marks a significant advancement in the description of magnetic systems.   Specifically, we theoretically identified a broad intermediate precritical regime B, situated between the low-temperature, spin-wave-dominated regime A and the critical ($T\Sb C - T \ll T\Sb C$)  regime C. In the  precritical regime, $M(T)$  follows a scaling behavior $M(T)\propto (T\Sb C -T)^{\beta_*}$, characterized by the precritical anomalous exponent $\beta_*\approx 1/3$. 

Our approach is not limited to simple ferromagnets but may be extended to ferrimagnetic materials with multiple magnetic sublattices, such as yttrium-iron-garnet, and antiferromagnets that involve exchange and dipole-dipole interactions, as well as anisotropy energy, among other factors.
  
\section*{Acknowledgments} 
We are grateful to E. Podivilov,  A. Shafarenko,  V. Cherepanov (now deceased) and V. Belinicher (now deceased)  for their contributions in the preliminary stages of this work many years ago. The authors also thank A. Chumak for the fruitful discussion. I.K. gratefully acknowledges the support of the  the support of the Russian Ministry of Science and High Education. 
V.L. was partially supported   through the project I-6568 “Paramagnonics.”
 \appendix
 \section{\label{s:DT}Diagrammatic technique for Heisenberg ferromagnets}
 \subsection{\label{ss:DT} Vaks-Larkin-Pikin DT for spin operators in thermodynamic equilibrium}
  
The initial formulation of the DT for ferromagnetic materials was proposed by Vaks, Larkin, and Pikin (VLP) \cite{Vaks1968a,Vaks1968b}. Their DT, formulated for the thermodynamic equilibrium, produced important results. 
Later Izyumov and Skryabin\, \cite{Izyumov1988}, and  Bar'yakhtar, Krivoruchko, and Yablonski\,\cite{Baryahtar1983}  adapted VLP DT for direct use with spin operators.
These DTs have produced several noteworthy and crucial findings  \cite{Vaks1968a,Vaks1968b,Izyumov1988,Baryahtar1983}. 

From the formal point of view, the  WH mean-field Eq.\eqref{QWEa} is the leading order approximation in these approaches, valid in the limit of an infinitely large radius of interaction $R$, defined as follows
\begin{equation}\label{B.2}
  R^2=\sum_{j} R_{i j}^2 J_{i j}/ a^2 J_0\ .
\end{equation}  
VLP also computed first-order corrections in $R^{-3}$ for $M(T,H)$ in \eqref{BriB} and the simultaneous spin correlation functions. According to \cite{Vaks1968a}, the expansion parameters of this theory for the cubic, body-centered (BCC), and FCC lattices with nearest-neighbor interactions are $1$, $2^{-3/2}\approx0.35$, and $3^{-3/2}\approx0.19$, respectively. However, since exchange interactions decay rapidly with distance, the theory, in its original formulation, is formally inappropriate for most ferrodielectrics.

 Most of the results obtained in this DTs \cite{Vaks1968a,Vaks1968b,Izyumov1988,Baryahtar1983} have not gained widespread acceptance, both because of the specific difficulties inherent in these diagram techniques and the unsuccessful graphical notation, which makes it difficult to establish analogies and perceive the information presented.

 Note also that in the VLP approach, the corrections to $M(T)$ become infinite as $T$ approaches $T_{_{C}}$. This behavior is not physical because $0 < M(T) < M_0$. Therefore, it is necessary to reformulate the theory to eliminate these infinite values. This is done below in Appendices \ref{ss:SP}, \ref{ss:FR} and 
 \ref{ss:1-loop}, where we provide a regularized theory that describes corrections to quantum WH-MFA that are applicable over a wider range of temperatures. 


\subsection{\label{ss:SP}Physical small parameter}

We have to note that from a physical point of view, MFA in ferromagnetics neglects the fluctuation of the effective magnetic field and becomes exact when the number of interacting magnetic atoms goes to infinity. For FCC crystals with nearest-neighbor interactions, this number is the coordinate number $Z_1=12$. Thus, the applicability parameter in this case is $1/Z_1= 1/12$,  which is much smaller than the formal expansion parameter $1/R^3=1$, declared by VLP. Moreover,  when the next-nearest interactions are important one has to account for their contribution to the  
exchange integral $J_0$. Thus we expect that in the general case, the role of coordinate number $Z$ is played by $Z\sb{eff}=J_0/J_1\approx Z_1+ Z_2 J_2/J_1$. In EuO $Z\sb{eff}\approx 13.2$ while in EuS $Z\sb{eff}\approx 9.4$. This explains why we hope to reach a better agreement between an experiment and the mean-field approach for EuO than for EuS. Indeed, according to Tab.\ref{t:1}, in  EuO $[T\sp{ex}\Sb{C}-T\sp{exp}\Sb{C}]/ T\sp{exp}\Sb{C}\approx 0.24$ while in EuS this ratio is about 0.29.   
  
\subsection{\label{ss:FR}Functional representation}
  
In this paper, we are interested in the spontaneous magnetization $M(T)$ which is proportional to the mean spin $\overline S_z$, defined by \eqref{meanS} and aligned with the external magnetic field  $\bm h=\{ 0\,, \ 0\,, h_z\equiv h\}$.
 Following \cite{Kolokolov1986,Kolokolov1989} we compute $M(T)$ as
 \begin{equation}\label{mean}
\overline{S}_z=\frac{ T }{ N_{\rm lat} }\frac{d}{dh}\ln {\cal Z}(h),\quad h\to 0\ .
\end{equation}

 from the generating function 
\begin{equation}\label{mFunk}
{\cal Z}(h)=\hbox{ Tr}\,\exp\Big( -\beta {\cal H}_{ex}+ \beta h\sum_j S^z_j\Big)\,.
\end{equation}

 Here, "Tr" represents the trace operator.
   The Heisenberg exchange Hamiltonian, denoted as  $ {\cal H}_{\rm ex} $, is derived from the exchange energy described in equation \eqref{EexB} and has the following form:
\begin{equation}\label{Hei-Hamiltonian}
{\cal H}_{\rm ex}=-\frac{1}{2}\sum_{jk}J _{jk}{\bm S}_j{\bm S}_k\,,
\end{equation}
where ${\bm S}_j$ are spin operators at the
lattice sites ${\bm R}_j$. 
To confine the operator problem to a single lattice site, the Hubbard-Stratonovich transformation\,\cite{Hubbard1959,Stratonovich1958} is utilized.
  
  The noncommutativity of spin components prevents us from rewriting the partition function as an integral over vector variables defined on each lattice site. Instead, the fields dependent on artificial time emerge, leading to an expression involving functional integrals over these fields defined on the lattice. 
  
First we represent ${\cal Z}(h)$ as an
infinite product:
\begin{align}\begin{split}\label{HubbS1}
{\cal Z}(h)=& \exp\Big(-\epsilon {\cal H}_{\rm ex}+
\epsilon h \sum_j S^z_j\Big)\times\dots \\
& \times
\exp\Big(-\epsilon
 {\cal H}_{\rm ex}+
\epsilon h \sum_j S^z_j\Big) \ .
\end{split}\end{align}
Formally, we need to let $\epsilon$ approach 0, but for now, let us consider it as a very small but finite quantity. The number of terms in the 
product (\ref{HubbS1}) is $\beta/\epsilon$. 
Then we use the identity valid in the limit $\epsilon\to 0$
\begin{align}\begin{split}\label{Gauss}
& \exp\left(\frac{\epsilon}{2}{ J}_{jk}{\bf S}_j{\bf S}_k
\right)  \\  
=& {\cal M}_\epsilon\int\prod_j d\mbox{\boldmath $\phi$}_j \exp\biggl(-\frac{\epsilon}{2}
{ J}_{jk}^{-1}\mbox{\boldmath $\phi$}_j\mbox{\boldmath
$\phi$}_k+\epsilon \mbox{\boldmath $\phi$}_j
{\bf S}_j\biggr),
\end{split}\end{align}
where ${ J}_{jk}^{-1}$ is the inverse matrix of ${ J}_{jk}$ and
${\cal M}_\epsilon$ is a normalization factor. 
Substituting   it into (\ref{HubbS1}), we have for each lattice site $j$ a set of integration variables corresponding 
to each multiplier in the product (\ref{HubbS1}). This set can be considered as a function ${\boldmath \phi}_j(t)$ defined in discrete
time $t=0, \epsilon, 2\epsilon, \dots \beta$. In a formal continuum limit, we  obtain  integral sums as an exponent and a product of 
time-ordered exponentials:
\begin{align} \nonumber \label{Fun1A}
{\cal  Z}(h)&=\int\prod_l {\cal D} \bm \phi _l(\tau)
 \exp\Big[-\frac{1}{2}\int_0^\beta
 { J}_{jk}^{-1} \bm \phi _j(\tau)  \bm \phi_k(t)\,d\tau  \Big]\\ 
 & \times \prod_j\mbox{ Tr}\,\mbox{T}  \exp\Big[\int_0^\beta d\tau \Big(
  \bm \phi_j(\tau)+{\bm h}_j\Big )
 {\bm S}_j\Big] \ .
 \end{align}
The symbol T denotes
a chronological product and ${\bm h}_j=(0,0,h)$.
The path integral (\ref{Fun1A}) is understood
as a limit of finite-dimensional approximations. The measure of integration is 
\begin{align} \label{Mes}
{\cal D}\mbox{\boldmath $\phi$}_l(t) = &{\cal M}
\prod_{\alpha=x,y,z} \prod_{n=1}^{\beta/\epsilon}
d\phi_l^{\alpha}(n\epsilon)\,, \ 
 \end{align}
  where $ {\cal M}= \left( 
  {\cal M}_\epsilon\right)^{\beta/\epsilon}$.  
  
Let us rewrite   \eqref{Fun1A} more conveniently by shifting the variables of functional integration $\bm \phi= \bm \varphi +\bm  h $: 
\begin{align} \label{Fun2}\nonumber  
{\cal Z}({\bm h})= & 
\! \!  \int\! \prod_l {\cal D} \bm \varphi _l(\tau)
\exp\Big [ -\frac{1}{2}\int_0^\beta
J _{jk}^{-1} \bm \varphi _j(\tau)
 \bm \varphi _k(\tau)d\tau \\  
 & +  \frac {h}{J_0}\int _0 ^\beta \sum _j
 \varphi ^z _j(\tau) d\tau - N_{\rm lat} \frac {\beta \,h^2}{2 J_0}\Big ] \\ \nonumber   &  \times
  \prod_j
\mbox{ Tr}\,\Big [ \mbox{T}\exp\biggl(\int_0^\beta
\bm \varphi _j(\tau) 
{\bm S}_j d \tau  \Big ]\ .
 \end{align} 
 
The time-ordered operator exponent
\begin{subequations}\label{Op}
\begin{equation}\label{OpA}
\hat{A}(t)=\mbox{T}\Big \{ \exp\Big [ \int_0^t
\bm \varphi (\tau)
{\bm S}d\tau\Big ]\Big \}
\end{equation}
is defined by the equation
\begin{equation}\label{Op-eq}
d\hat{A}(t)/dt=\Big[ \bm \varphi (t)
{\bm S}\Big ] \hat{A}(t)
\end{equation}
\end{subequations}
with the initial condition $\hat{A}(0)=1$. The operator $\hat{A}$
cannot be expressed explicitly as a functional of
$\bm \varphi (t)$. However, there is a way to rewrite the T-ordered exponential as a product of regular ones.   To demonstrate this, let us consider the operator given as a product of the usual matrix exponential: 
\begin{align}\begin{split}\label{Anzatzm}
 \widehat{\cal A}   & (t) 
 =  \exp \left[-  S^+\psi^-(t)\right ]
\exp \Big[ S^z\int\limits_{0}^{t} \rho(\tau)\,d\tau\Big ] \\
&     \times \exp \Big \{ \frac 12\,  S^-  \!  \int\limits_{0}^{t}\,d\tau\psi^+(\tau)\exp\Big [ 
\int \limits_{0}^{\tau} \rho(\tau^\prime)\,d\tau^\prime\Big ]
\Big \} \\ 
& \times \exp \Big [ S^+\psi^-(0)\Big ]\ .
\end{split}\end{align}
Here, $S^{\pm}=S^x\pm iS^y$ and $\psi^{\pm}(t), \rho(t)$ are some new fields.
Using the commutators
\begin{align}\begin{split}\label{comm}
[S^-, F(S^+)]=& -2S^zF^\prime (S^+)+S^+F^{\prime\prime} (S^+)\,, \\
\ [S^z, F(S^-)]=&-S^-F^{\prime}(S^-)\,,\\
 [S^+, F(S^-)]=& 2S^zF^\prime (S^-)+S^-F^{\prime\prime} (S^-)\,, \\
 S^-F(S^z-1)  =& F(S^z)S^-\,, 
\end{split}\end{align}
where $F(y)$ is some function, one can be convinced that the operator
$\hat{{\cal A}}(t)$ satisfies  the following equation:  
\begin{align}\begin{split}\label{Anzeq1}
\frac{ d\widehat{\cal A}} {d\tau}=& \Big \{ \big[\rho - \psi^+\psi^-\big ] S^z+
\frac 12 \psi^+S^- \\ & + \big [-\dot{\psi^-} +    \rho\psi^- + \frac 12 \,\psi^+(\psi^-)^2 \big ] S^+\Big \}
\hat{\cal A}(t).
\end{split}\end{align}
The last factor in (\ref{Anzatzm}) provides the equality $\hat{{\cal A}}(0)=1$.
 Considering the following change of variables in the functional integral over the
measure  
\begin{align}\begin{split}\label{Changem}
\varphi^z=& \rho-\psi^+\psi^-\,, \\
 \varphi^- =& - \dot{\psi^-} + \rho\psi^- -  \frac 12\, \psi^+(\psi^-)^2 \, ,\\
 \varphi^+=&\frac 12 \, \psi^+ \,,\quad \varphi^{\pm}=\frac 12 \,(\varphi^x\pm i\varphi^y) \,,
 \end{split}
 \end{align}
we see that   the T-ordered operator  $\hat{A}(t)$, defined by Eqs.\,\eqref{Op}, takes   the form (\ref{Anzatzm}):
\begin{equation}\label{mAnzz}
\widehat{A}(t)=\widehat{\cal A}(t)\ .
\end{equation}
without T-ordering. 
This allows us to obtain an explicit  functional integral
representation for any physical quantities of interest.

The change of variables (\ref{Changem}) contains the time derivative of $\psi^-$ on the right-hand side. Hence, it is necessary to impose the boundary or initial conditions.

Here, we utilize the periodic boundary conditions typically used in the statistical physics of Bose systems.
It is essential to calculate the Jacobian $\cal J$ considering the boundary conditions:
 \begin{equation}\label{mJacrez}
  {\cal J}=
\mbox{const}\,  \sinh \Big  (\frac 12 \,\int\limits_{0}^{\beta}
\rho\,dt \Big ) \ .
\end{equation}
The analyticity of the integrand and the convergence of the
functional integral allows us to deform the initial surface of
integration into a standard one: $\hbox{Im} \rho=0$,
$\psi^+=(\psi^-)^*$.
In this way, the trace of the operator $\hat{{\cal A}}(\beta)$ (\ref{Anzatzm})
can be easily calculated for an
arbitrary value of the spin $S$: 
\begin{equation}\label{Trace}
\mbox{Tr}[\widehat{\cal A}(\beta)]=
\mbox{Tr}[\hat{A}(\beta)]=\exp\Big[{\cal Q}_{_S} \Big(\int\limits_{0}^{\beta}
\rho\,dt \Big)\Big]\ .
\end{equation} 
Here ${\cal Q}_{_S}(x)$ is a primitive of normalized Brillouin function $b_{_S}$, givn by \eqref{bS}:  
\begin{subequations}\label{pBrill}
\begin{align}  \label{pBrillA}
{\cal Q}_{_S}(x)=& \ln \frac{\sinh\left[x(S+1/2)\right]}{\sinh(x/2)} \\
b_s(x)=& d{\cal Q}_{_S} (x)/dx\ .  
 \end{align}\end{subequations}
Note that ${\cal Q}(x)$ differs from the original Brillouin function ${\cal B}_{_S}(x)$, given by \eqref{BriA}.
 
 Thus, according to \eqref{mean}, \eqref{Fun2}, and \eqref{Changem}-\eqref{pBrill}, the spontaneous magnetization is given by  the expectation value 
\begin{align}\begin{split}\label{MeanS}
\bar{S}=&  \frac{T}{N_{\rm lat} J_0}\Big\langle \int_0^\beta
\sum_j\varphi^z_j(\tau)d\tau \Big\rangle\\  =& \frac{T}{N_{\rm lat} J_0}\Big\langle\int_0^\beta
\sum_j \left(\rho_j - \psi^+_j\psi^-_j\right)(\tau)d\tau \Big\rangle
\end{split}\end{align}
with respect to the measure 
\begin{equation}\label{merra}
{\cal D}\rho {\cal D}\psi^+{\cal D}\psi^- 
\exp\Big[- \int_0^\beta {\cal L}\,d\tau +\sum_j g_s\Big(\int_0^\beta \rho_j\, d\tau\Big)\Big],
\end{equation}
where the Lagrangian $ {\cal L}(\rho,\psi^\pm)$ has the form:
\begin{align}\begin{split}
{\cal L}(\rho,\psi^\pm)= & \sum_{j,l}\Big[\frac{1}{2}\rho_j{\cal J}_{jl}^{-1}\rho_l  -
\psi^+_j{\cal J}_{jl}^{-1}\dot{\psi}^-_l  \\ & +
\rho_j{\cal J}_{jl}^{-1}(\psi^-_j \psi^+_l - \psi^-_l \psi^+_l)  \\ 
&+ \frac{1}{2}\psi^-_j \psi^+_j{\cal J}_{jl}^{-1}\psi^-_l \psi^+_l \\
&-\frac{1}{2}\psi^-_j {\cal J}_{jl}^{-1}(\psi^-_l)^2 \psi^+_l \Big] \,, 
\end{split}\end{align}
 with ${\cal J}^{-1} $ being the inverse Jacobian \eqref{mJacrez} and the function $g\Sb S(x)$ is 
  \begin{subequations}
\begin{align}\begin{split}\label{gS}
  g_{_{\rm S}}(x)= & {\cal Q}_S(x)+\ln\Big [\sinh\Big (\int\limits_{0}^{\beta}
\frac{1}{2}\rho\,d\tau \Big )\Big ]\\ =& \ln\Big [\exp\Big ((S+\frac{1}{2})x\Big )\\
&-\exp\Big (-(S+\frac{1}{2})x\Big )\Big ]\,, \end{split}\\ 
 \begin{split}\label{gS1}
\frac{d}{dx}g_{_S}(x)=&\frac{1}{2}+b_S(x)+n_0(x)\, ,\\ n_0(x)=&\frac{1}{e^x -1} \ .
 \end{split}
 \end{align}
 \end{subequations}

\begin{figure*}   
 \includegraphics[width=1.8\columnwidth]{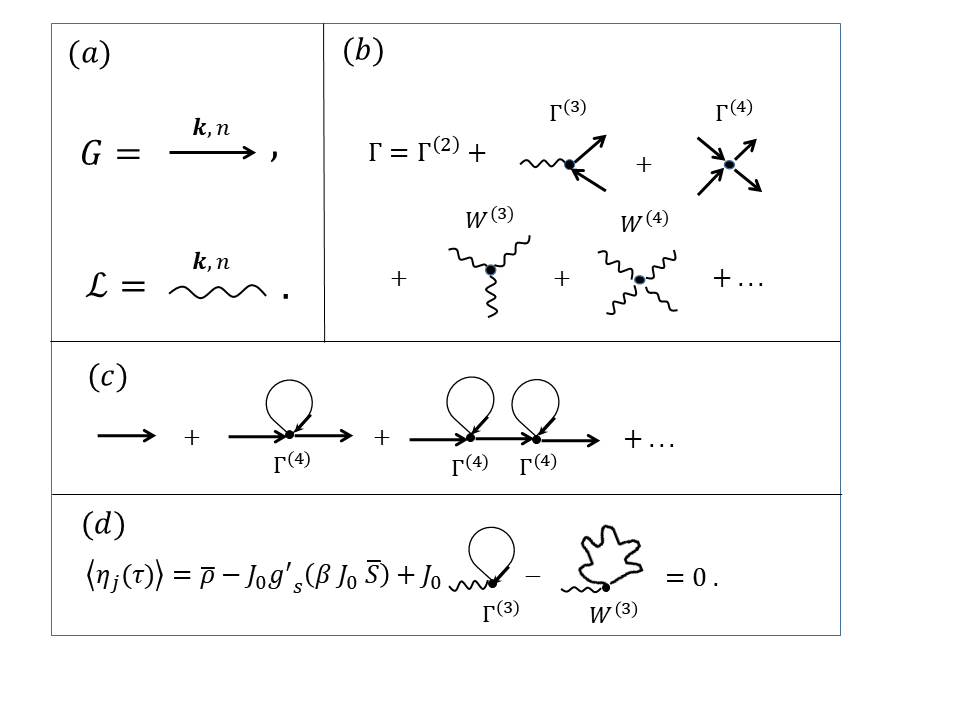} 
 \caption{\label{f:4}  (a) Graphical notation for the correlators  $G_0 (\omega_n, \bm k)$, and  ${\cal L}_0(\omega_n, \bm k)$.  (b) notation for vertices $\Gamma^{(3)}$, and $\Gamma^{(4)}$, $ W^{(3) }$, and $W^{(4)}$.
 (c) One-loop renormalization of the frequency spectrum. (d) The one-loop equation for magnetization. }
 \end{figure*}

\subsection{\label{ss:1-loop}One-loop equation for spontaneous magnetization}
\subsubsection{\label{sss:Measure}Integration measure  in the perturbation approach}
To prepare the integration measure over fields $\rho_j,\,\psi_j^\pm$ at the site $\bm r_j$, we note that in our system with a nonzero mean spin $\overline S_z$, the field $\rho_j(t)$ also has a non-zero mean value $\overline \rho$:
\begin{equation}\label{I-1.1}
\rho(\bm r_j,t) \equiv \rho_j(t) =\overline \rho + \eta_j(t)\,,
\end{equation}
where fluctuations $\eta_j(\bm r_j,t)\equiv \eta_j(t)$ are assumed to be small in some sense.

In our case of spatial and time homogeneity, it is customary  to use Fourier components 
\begin{subequations}\label{I-1.2}
\begin{align}\label{I-1.2A}
 \psi^\pm_j( t)= & \frac 1 N_{\rm lat} \sum_{n,\bm k}\psi^\pm _{n, \bm k}\exp [\pm i (\omega_n  \tau  + \bm k \cdot \bm r_j)]\,, \\ \label{I-1.2B}
\eta  _j( t)= & \frac 1 N_{\rm lat}\sum_{n,\bm k}\eta _{n, \bm k}\exp [  i (\omega_n \tau  + \bm k \cdot \bm r_j)] \ .
\end{align}
\end{subequations} 

 The equation for  spontaneous magnetization follows from the identity
 \begin{equation}\label{idde}
\langle \eta_j(\tau) \rangle =0
\end{equation}
valid for any $j$ and $\tau$ due to the homogeneity. To compute this expectation value explicitly, we substitute the decomposition  \eqref{I-1.2}   into 
(\ref{merra})-(\ref{gS}) and arrive at the measure of perturbative averaging over fluctuations around the mean  field:

 \begin{subequations}\label{I-1.4}
\begin{align}\label{I-1.4A}
&{\cal D}\eta {\cal D}\psi^\pm \exp (- \Gamma)\,, \quad \mbox{where}\\ \label{I-1.4B}
\Gamma =&  \Gamma^{(2)}+ \Gamma^{(3)}+ \Gamma^{(4)}- W\{\eta \}\,, \\
 \begin{split}\label{I-1.7}
\Gamma^{(2)}=& \sum_{n, \bm k}\frac1 {J_{\bm k}} \Big \{ \frac{|\eta_{n,\bm k}|^2 }2   \Big [  1 - \beta^2 J_{\bm k} g_{_S} '' (\beta \overline \rho) \Delta (n)\Big]\\
&+ \psi^+_{ n  ,\bm k} \psi^-_{ n ,\bm k}\Big [i \omega _n + \overline \rho \Big ( 1- \frac {J_{\bm k}}{J_0}\Big) \Big ] \Big \} \ .
\end{split}\end{align}
Hereafter $\Delta(n)$  is the Kronecker delta function defined as $\Delta(0)=1$, and   $\Delta(n\ne 0)=0$. Furthermore,
  \begin{align} \label{I-1.8A}
& \Gamma^{(3)}=\frac 1 {\sqrt N\sb{lat}} \sum_{n_{1,2,3}, \bm k_{1,2,3}} \eta_{\bm k_1, n_1}\psi^+_{\bm k_2, n_2}\psi^-_{\bm k_3, n_3} \\ \nonumber
 \times &  Q_{\bm k_1,\bm k_2}  \Delta (n_1+n_2-n_3)\Delta (\bm k_1+\bm k_2-\bm k_3)\,,\\
& Q_{\bm k_1,\bm k_2 }=  \frac 1 {J_{\bm k_2}}- \frac 1 {J_{\bm k_1}}\,, \\
 \nonumber 
& \Gamma^{(4)}=\frac 1 {\sqrt N\sb{lat}} \sum_{n_{1,2,3,4},  \bm k_{1,2,3,4}}T_{\bm k_{1,2};\bm k_{3,4}}\\ \label{I-1.9A}
& \times  \psi^+_{\bm k_1, n_1}\psi^+_{\bm k_2, n_2}\psi^-_{\bm k_3, n_3} \psi^-_{\bm k_4, n_4} \\ \nonumber
  & \times  \Delta (n_1+n_2-n_3-n_4)\Delta (\bm k_1+\bm k_2-\bm k_3-\bm k_4)\,, \\ \label{I-1.9B}
  &  T_{\bm k_{1,2};\bm k_{3,4}} =   \frac 14 \Big (\frac 1{J_{\bm k_1-\bm k_2}}+  \frac 1 {J_{\bm k_2-\bm k_4}}  -\frac 1{J_{\bm k_1}} + -\frac 1 {J_{\bm k_2} } \Big ) \ . 
 \end{align}
 \end{subequations}
Vertex $T_{\bm{k}_{1,2};\bm{k}_{3,4}}$ is independent of Matsubara frequencies $\omega_n$ and, due to the conservation law of momentum, is symmetric with respect to the permutation $\bm{k}_3 \leftrightarrow \bm{k}_4$.

Last but not least, the contribution to $\Gamma$ in \eqref{I-1.4B}, denoted as $W\{\eta \}$ , is an infinite series over fluctuations $\eta_{\bm k}\equiv \eta_{n=0, \bm k}$ with zero Matsubara frequencies:
 \begin{subequations}\label{I-1.12}
\begin{align}\nonumber
W\{\eta \}=& \sum _m N^{(1-m/2)} \frac {\beta^m}{m!}\sum _{\bm k_1 ... \bm k_m}\Delta(\bm k_1 +... +\bm k_m)\\ \label{I-1.12A}
&\times g_{_S}^{(m)}(\beta \overline \rho)\eta_{\bm k_1}\dots \eta_{\bm k_m}\,,\\ \label{I-1.12B}
g_{_S}^{(m)}(x)=& \Big ( \frac d {d\, x}\Big )^mg_{_S} (x)\ .
\end{align}\end{subequations}


\subsubsection{\label{sss:cf} Magnetization in the first order in $1/Z$}
Seed correlation functions are diagonal in $\bm k$ and $\omega_n$:
 \begin{align} 
 \begin{split} 
G_0 (\omega_n, \bm k)=& \langle \psi^+ _{n,\bm k}  \psi^+ _{n,\bm k}\rangle \\ 
=&   J_{\bm k}\Big [ i \omega _n + \frac {(J_0 - J_{\bm k})\overline \rho}{J_0}\Big ]^{-1}\,,
  \\ \label{I-1.17}
  {\cal L}_0(\omega_n, \bm k)= & \langle \eta _{n,\bm k}  \eta _{-n,-\bm k}\rangle \\ =& \frac {J_{\bm k}}{1- \beta^2J_{\bm k} g_S'' (\beta \overline \rho)\Delta(n)} \ .
 \end{split} \end{align} 
 Graphical notations for correlators $G_0 (\omega_n, \bm k)$ and  ${\cal L}_0(\omega_n, \bm k) $ are shown in Fig.\,\ref{f:4} together with graphical notations for vertexes $\Gamma^{(3)}$, $\Gamma^{(4)}$, and $W$, defined by Eqs.\,\eqref{I-1.8A}, \eqref{I-1.9A} and  \eqref{I-1.12}.

The one-loop renormalization of the Green function $G$ depicted in Fig.\,\ref{f:4}(c)  
is equivalent to substitution $\bar{\rho}\to J_0\bar{S}$ in the expression  \eqref{I-1.17} for $G_0$. 
Additionally, the regularization of simultaneous field products related to the initial spin problem follows the Stratonovich rule \cite{Stratonovich1958} rather than relying on chronological ordering.
It is equivalent to a symmetrical $n\to -n$ cut-off in summation over Matsubara frequencies. It gives
\begin{equation}\label{psipsi}
\langle \psi^-_{\bm k} \psi^+_{\bm k}\rangle = J_{\bm k}\left(\frac{1}{2}+n_{\bm k}\right), \quad
n_{\bm k}=\frac{1}{e^{\beta E_{\bm k}} -1}.
\end{equation}
It can be verified that equation \eqref{psipsi}, along with Jacobian \eqref{mJacrez}, is consistent with the kinematic identities \eqref{KI} described in Section \ref{ss:BL-DT}. For example, in the case where $J_{nn'} \propto \delta_{nn'}$, the free energy is proportional to $ S(S+1) $.

Note that Eq.\,\eqref{idde} for $\langle \eta \rangle$ represents the sum of connected diagrams. In the one-loop approximation, we have the terms shown in Fig.\,\ref{f:4}(d). By substituting the analytical expressions for the propagators and vertices provided above, we arrive at Eq.\,\eqref{I1.22A}.

%


\end{document}